\documentclass [12pt,twoside]{article}           % version LATEX2E

\usepackage{epsfig,times,lscape}
\usepackage[usenames]{color}

    \ifnum\count102=1

    \fi

    \ifnum\count102=2
\fi

\newcommand{\nc}{\newcommand}

     \ifnum\count101=1
\nc{\qI}[1]{\section{{#1}}}
\nc{\qA}[1]{\subsection{{#1}}}
\nc{\qun}[1]{\subsubsection{{#1}}}
\nc{\qa}[1]{\paragraph{{#1}}}

\def\qpar{\vskip 2mm plus 0.2mm minus 0.2mm}
\def\qL{\hfill \break}
     \fi 

      \ifnum\count101=2
 \nc{\qI}[1]{\parindent=0mm \vskip 8mm 
{\centerline{\LARGE \color{red}#1}}\vskip 3mm}
\nc{\qA}[1]{\vskip 2.5mm \noindent 
{{\bf\large\color{blue}  #1}} \vskip 1mm \parindent=0mm}
 \nc{\qun}[1]{\vskip 1mm \noindent {\sl #1 }\quad }

\def\qL{\hfill \break}
\def\qpar{\vskip 2mm plus 0.2mm minus 0.2mm}

      \fi
\def\qth{\vrule height 12pt depth 0pt width 0pt}
\def\qtb{\vrule height 0pt depth 5pt width 0pt}

\nc{\qfoot}[1]{\footnote{{#1}}}

      \ifnum\count101=1
\def\qbu{\hfill \par \hskip 6mm $ \bullet $ \hskip 2mm}
\def\qee#1{\hfill \par \hskip 6mm (#1) \hskip 2 mm}
      \fi
      \ifnum\count101=2
\def\qbu{\hfill \par \hskip 4mm $ \bullet $ \hskip 2mm}
\def\qee#1{\hfill \par \hskip 4mm (#1) \hskip 2 mm}
      \fi

\def\qparr{ \vskip 1.0mm plus 0.2mm minus 0.2mm \hangindent=10mm
\hangafter=1}

     \ifnum\count101=1 
 
     \fi
     \ifnum\count101=2

  \def\qcitb#1{\noindent \hbox to 102mm{\hfill \small #1} \vskip 1mm}
      \fi

 \def\qpages#1{\count102=0{\loop\advance\count102 by 1
 \null \vfill\eject \ifnum\count102<#1 \repeat}}

\def\qn#1{\eqno \hbox{(#1)}}

\def\qth{\vrule height 12pt depth 0pt width 0pt}
\def\qtb{\vrule height 0pt depth 5pt width 0pt}

\def\qv{\vskip 0.1mm plus 0.05mm minus 0.05mm}
\def\qhu{\hskip 0.6mm}
\def\qhv{\hskip 3mm}

\def\qhw{\hskip 1.5mm}
\def\qleg#1#2#3{\noindent {\bf \small #1\qhw}{\small #2\qhw}{\it \small #3}\qv }

\begin{document}
\thispagestyle{empty}
% --------------------------------------------------------------------

      % Hauts de pages et numerotation

          % Remarque: sans le \protect --> message d'erreur (ordre fragile)
\markboth{{\sl \hfill  \hfill \protect\phantom{3}}}
        {{\protect\phantom{3}\sl \hfill  \hfill}}

% -------------------------------------------------------------------
\color{yellow} 
\hrule height 20mm depth 10mm width 170mm 
\color{black}
\vskip -2.5cm 
\centerline{\bf \Large How does group interaction and its severance}
\vskip 2mm
\centerline{\bf \Large affect life expectancy?}
\vskip 4mm

\centerline{\large 
Lei Wang$ ^1 $,
Yijuan Xu$ ^1 $,
Zengru Di$ ^2 $,
Bertrand M. Roehner$ ^3 $
}

\vskip 15mm
\normalsize

{\bf Abstract}\quad The phenomenon of apoptosis observed
in cell cultures consists in the fact that  unless cells
permanently receive a ``Stay alive'' signal from their
neighbors, they are bound to die. A natural question
is whether manifestations of this apoptosis paradigm can
also be observed in other organizations of living organisms.
In this paper
we report results from a two-year long campaign of
experiments on three species of ants and one species of
(tephritid) fruit flies. 
In these experiments individuals were separated
from their colony and kept in isolation either alone
or in groups of 10 individuals. 
The overall conclusion is that ``singles''
have a shorter life expectancy than individuals in
the groups of 10. This observation holds for ants
as well as for fruit flies. The paper also provides
compelling evidence of a similar effect in
married versus unmarried (i.e. single, widowed or divorced) 
people. 
A natural question concerns the dynamic of the transition
between the two regimes. Observation suggests an abrupt (rather
than smooth) transition and this conclusion seems to hold
for ants, fruit flies and humans as well. We call it a
shock transition. 
In addition, for red fire ants {\it Solenopsis invicta},
it was observed that individuals
in groups of 10 that also comprise one queen, die much faster
than those in similar groups without queens.\qL
The paper also examines the corresponding
survivorship curves from the
perspective of the standard classification into 3 types.
The survivorship curves of ants (whether single or in groups of 10)
are found to be of type II whereas those of the fruit fly
{\it Bactrocera dorsalis} are rather of type III.
In this connection it is recalled that the survivorship curve
of the fruit fly {\it Drosophila melanogaster} is
of type I, i.e. of same type as for humans. 

\vskip 6mm
\centerline{\it First version: 28 March 2013, comments are welcome}

\vskip 6mm
{\normalsize Key-words: Group effect, collective behavior,
living organisms, insects, ants, drosophila, marital status,
widowhood, social bonds, coupling strength, survivorship curve.}
\vskip 2mm

PACS classification: Interdisciplinary (89.20.-a) +
collective effects (71.45.Gm)
\vskip 8mm

{\normalsize 
1: Laboratory of Insect Ecology, 
Red Imported Fire Ants Research Center, South China Agricultural
University, Guangzhou, China.\qL
Emails: wanglei\_1107@yahoo.com.cn, xuyijuan@scau.edu.cn
\qL
2: Department of Systems Science, Beijing Normal University, 
Beijing, China.\qL
Email: zdi@bnu.edu.cn 
\qL
3: Institute for Theoretical and High Energy Physics (LPTHE),
University Pierre and Marie Curie, Paris, France. \qL
Email: roehner@lpthe.jussieu.fr
}

\vfill\eject

\large

To many readers this paper will appear somewhat unconventional.
There is a simple reason for that. It results from a 
collaboration between physicists and entomologists.
Physicists always try to find fairly general rules.
Of course, all rules and laws have limitations, but the 
broader the better%
\qfoot{As an illustration one can recall that the
law of free fall discovered by Galileo
four centuries ago applies to balls of metal as well as
to apples, nuts or hailstones and many other objects falling 
in air. Yet, it does {\it not} apply
to falls occurring in water instead of air.
Even in air the law of free fall no longer holds
for small objects of a size under one micrometer.}%
.
This is why this paper does not limit itself to just
one specific species but rather offers a comparative approach
involving several species. This comparative perspective is
one of the unconventional features we were referring to.
\qpar
A second characteristic of this paper
can be explained as follows. Physics
experiments are more than just careful observations; they
always ask Nature specific questions. If an experiment is well
designed%
\qfoot{Which in particular means that by choosing appropriate
experimental conditions the ratio signal/noise 
should be made as high as possible.}
Nature will provide a clear answer. 
The present study was set up in a
similar way. The question that we submit to Nature is the following.\qL
\centerline{\it Is the lifespan of living organisms affected by 
changes in group interaction?}
\qpar

This is not a new question. As a matter of fact, we will see
that an attempt to get an answer was made as early as 1944.
Yet, to our best knowledge, in the past decades there have been
few investigations focusing
on this question from a broad comparative perspective.
\qpar
% Limiter le bruit en amplifiant l'effet
There is a third noteworthy characteristics of our 
experimental methodology but we will postpone its 
discussion until the end of the paper.
\qpar

The paper is organized as follows. First, we
discuss relevant results published in former decades
and at the same time we explain how we came to study 
this question.
Secondly, we present our experiments. Their
methodology will be discussed and detailed
results will be given mostly in graphical form.
In the following part, we try to make some sense of our
results by considering them in a broad systems science
perspective. Finally, in the conclusion, we summarize our results
and we discuss 
a possible agenda for further investigations.

\qI{How interaction-changes affect survival: former studies}

The present questioning started in September 2005 when we came across
a paper describing an experiment
done in 1944 by two eminent French entomologists, 
Pierre-Paul Grass\'e and R\'emy Chauvin%
\qfoot{Whereas Grass\'e was already well-known,
Chauvin who was almost 20 years younger became famous only
in subsequent decades. He gave a great impulse to the study
of the collective behavior of insects. Unfortunately,
due to the invasive development of genetics and molecular
biology, this field was progressively deserted which explains that
after his retirement Chauvin had no followers whatsoever.}%
.
They separated small groups of social insects (of 1 to 10
individuals) from their colony in order to measure their
respective life span. The groups of 10 individuals turned
out to have a longer life expectancy that the groups of 1 or 2.
The corresponding graphs are given below in Fig. 1.

%
%%%% GRASSE ET CHAUVIN
\begin{figure}[htb]
\centerline{\psfig{width=12cm,figure=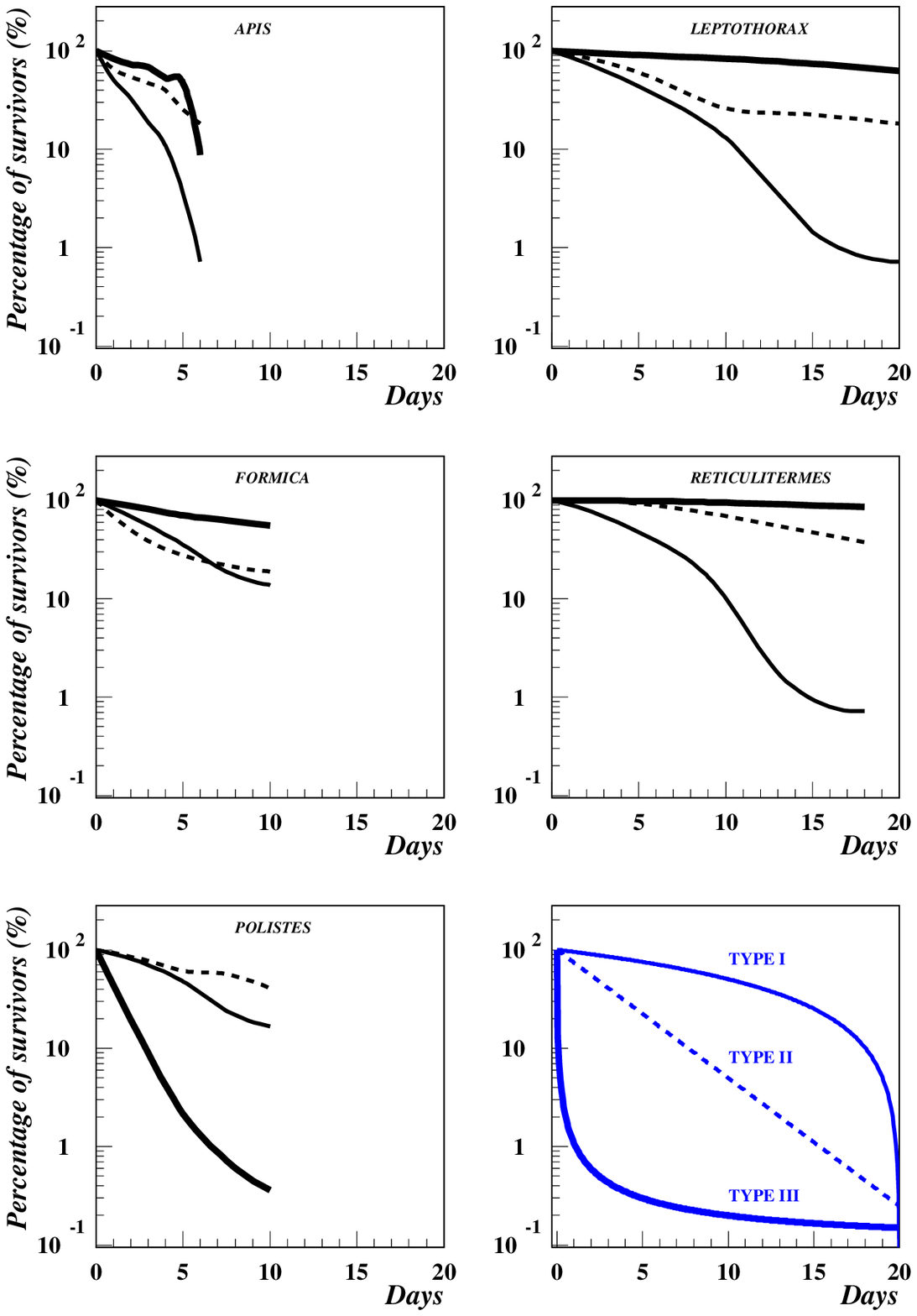}}
\qleg{Fig.\qhu 1\qhv Survivorship curves of social insects removed
from their colony.}
{Several groups of 1, 2 and 10 insects were removed from their
nest and kept in isolation. 
They received plenty of food
and were kept in conditions which matched as closely as possible those
in the nest.
The vertical scale shows the cumulative percentage of survivors in
each group. The thin solid lines represent single insects, the broken
lines represent groups of two insects and the thick solid line
represents groups of 10 insects. 
Initiated in 1944 by P.-P. Grass\'e and R. Chauvin this kind
of experiment may  provide a methodology for gauging the 
strength of the interactions in the colony. 
{\it Apis} designates a species of bees, {\it Leptothorax} and
{\it Formica} are two species of ants, {\it Reticulitermes} are termites and
{\it Polistes} are wasps. It can be seen that all these survivorship
curves are (approximately) of type II, except those for bees which
are rather of type 1 although in this later case the bees die
so quickly that the curve is not well defined.
The last graph shows the three standard survivorship curves.
Type I corresponds to the case of humans (and also drosophila); it
is characterized by low death rate for most of the life until
an age around 70.}
{Source: Grass\'e and Chauvin (1944); Roehner (2007, p. 236); see also
Arnold (1978).}
\end{figure}
%----------------------------------------------
%

Of the 3 types of survivorship curves shown in Fig. 1,
the simplest one is Type II because it corresponds to 
a mortality rate that is constant over time and
independent of age. Type III corresponds to a very high
``infant'' mortality (e.g. when thousands of seeds are released
of which only a few will grow). In contrast, Type I corresponds
to low infant mortality and a mortality rate that becomes
substantial only in old age. 
\qpar
Apart from the 3 types of survivorship curves shown in Fig. 1,
there is a 4th type namely the so-called bathtub curve. In this
case, both the infant mortality and the old age mortality
are high and are separated by a  range in which the mortality
is low and fairly constant (the ``bathtub'' label originates
from this shape).
The resultant survivorship
curve looks very much like Type III with the addition of
a downward section in old age. The bathtub curve applies mainly to 
technical devices such as ball bearings, electronic components, relays
or hard drives.
\qpar

Of course, the idea that the lifespan of individuals may
be affected by their interaction with the group to
which they belong was not completely new. Surprisingly, however, 
very little reliable experimental data seemed to be available
on this issue.
The results obtained by Grass\'e and Chauvin were 
a good starting point because they were clear, straightforward and
covered different species. 

\qpar
Then, in 2008 we discovered similar experiments done for cells
by Prof. Martin Raff and his collaborators (Ishizaki 1993,1994).
%%
%%%% RAFF
\begin{figure}[htb]
\centerline{\psfig{width=10cm,figure=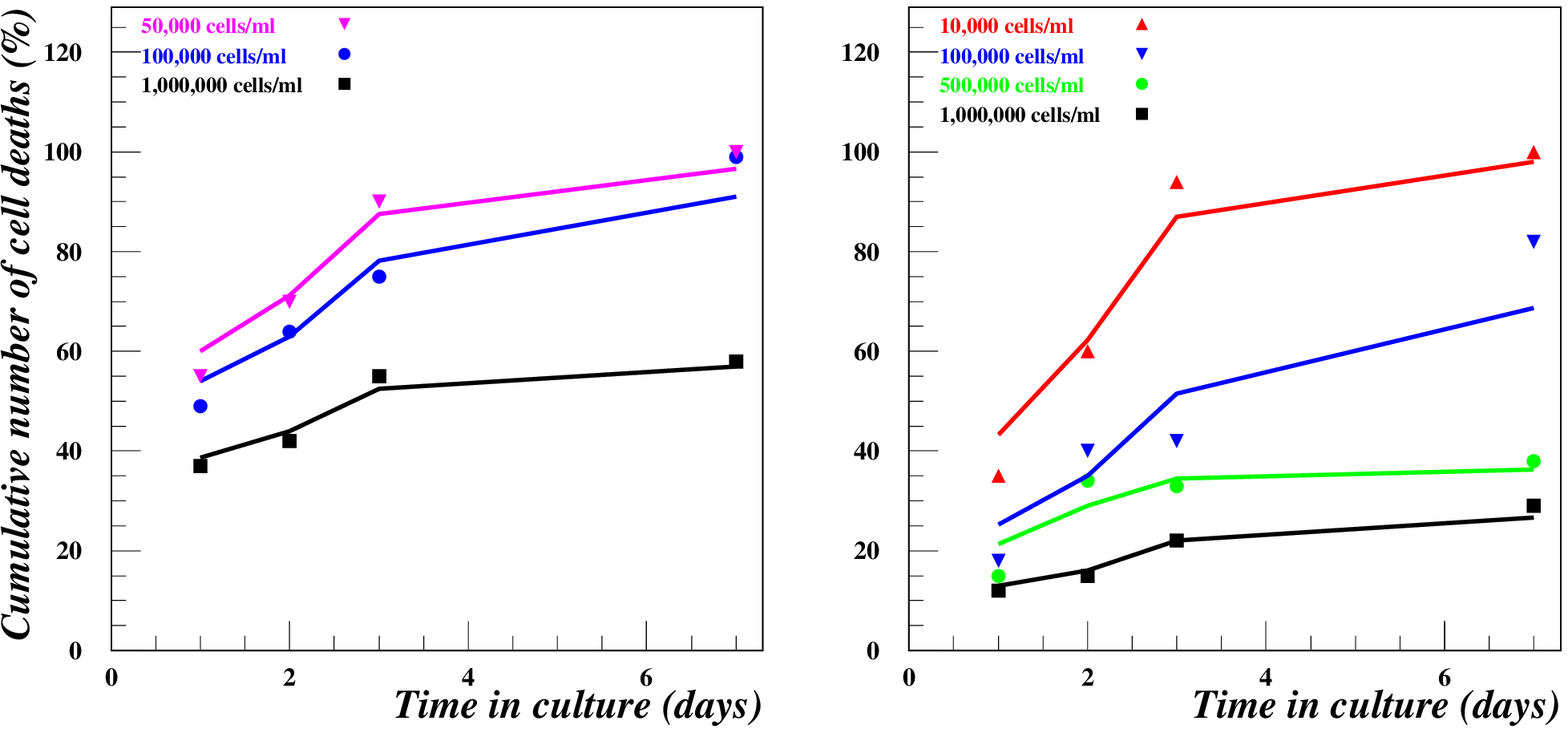}}
\qleg{Fig.\qhu 2\qhv Relationship between cell-density in culture and
their death rate.}
{Left-hand side: cells from the surface (epithelium)
of the crystalline lens 
of rat. Right-hand side: cells from the cartilage of rat.
It can be seen that the higher the density, the lower is 
the death rate. 
In both cases the culture was made on a gel of agarose which means
that the cells were disjoint and could not communicate by contact.}
{Source: Adapted from Ishizaki et al. (1993 p. 904, 1994 p. 1072).}
\end{figure}
%-------------------------------------------------
%
\qpar
For two different types of
cells, they showed that their {\it in vitro} lifespans increased 
with their density (see the graphs in Fig. 2).
Prof. Raff summarized this result (along with similar ones)
by saying that the cells will die 
unless they get from their neighbors 
repeated signals saying ``Stay alive, stay alive''.
These experiments were done in 1993 and 1994.
Yet, according to Prof. Raff (in an email of 4 August 2008)
few (if any) similar investigations were conducted 
in the following years. This means that we still do not know
what is the field of validity of this mechanism. Is it
common to most cells? Is it also shared by bacteria?
\qpar

\qI{Mortality according to marital status}

Around the same time, thanks to 19th century data and data 
from US yearbooks, 
we became aware of the fact that the death
rates of non-married persons (i.e. persons
persons who are single, widowed or divorced) are on average
about two times higher than the death rates of married persons.
(see Fig. 3a,b).

\qA{Bonds between husbands and wives}

Before giving fairly recent data we
first provide evidence from the 19th century and for different 
countries. Our purpose is to show that the rule holds
not just for one country in a specific times but under
a fairly broad range of conditions. Fig. 3a shows
the ratios of mortality rates of non-married to married
people in France, Belgium and the Netherlands.

%%
%%%% MORTALITE PAR ETAT MATRIMONIAL: BERTILLON
\begin{figure}[htb]
\centerline{\psfig{width=16cm,figure=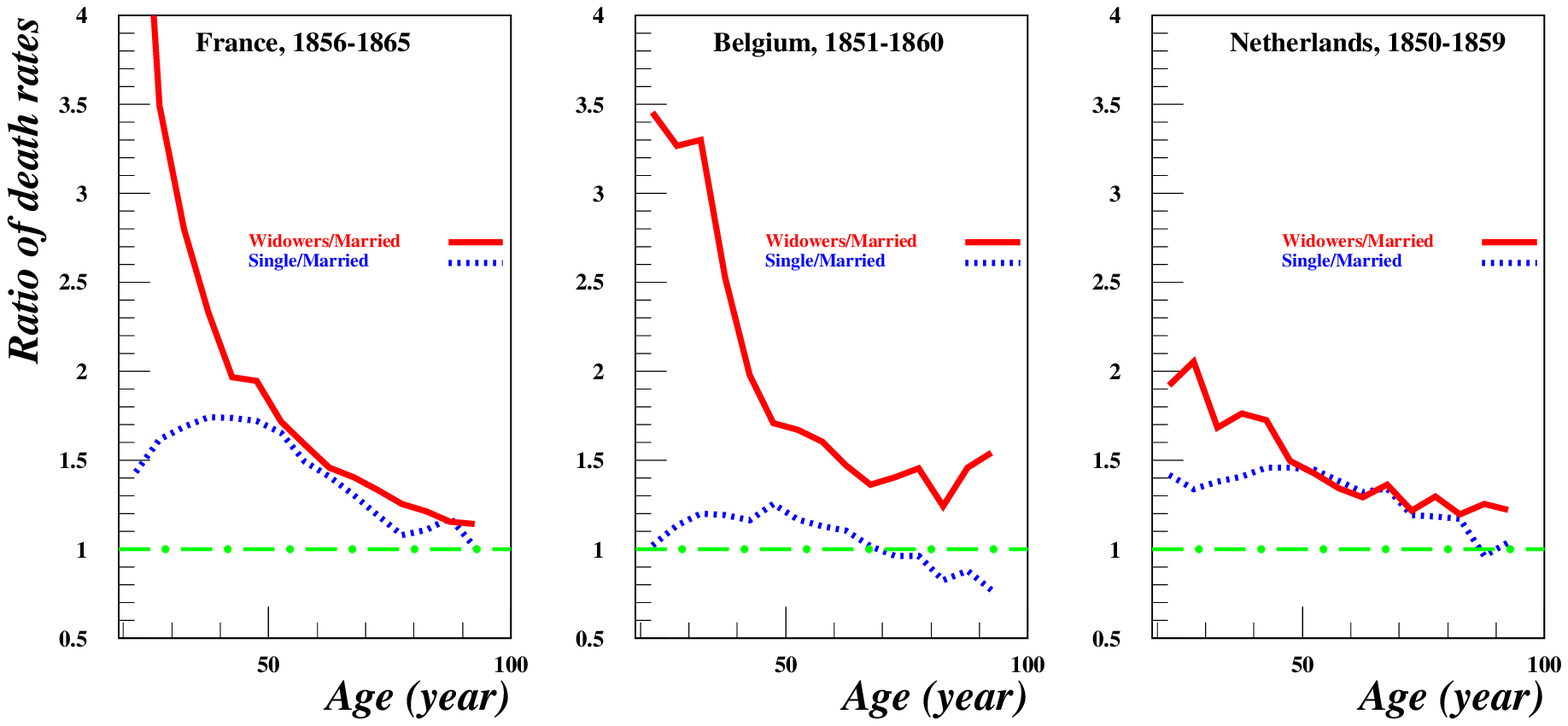}}
\qleg{Fig. \qhu 3a\qhv Ratio of mortality rates 
according to marital status.}
{The graph is based on mortality rates for 5-year age groups;
altogether, from 20 to 95 there are 15 age groups.
As a rule, the death rate of married persons appears
to be lower than
the death rate of non-married persons. The ratio is
larger for widowers than for bachelors. The results
for females are similar but the effect is weaker.
The primary data compiled by Bertillon give death rates 
of Belgian bachelors above the age of 80
that are on average 1.7 times smaller
than corresponding rates in France and the Netherlands;
this makes one suspect a possible statistical bias
in the Belgian data for this late age interval.}
{Source: Bertillon (1872)}
\end{figure}
%-------------------------------------------------
%
The fact that the ratio widowers/married is higher for
young people than in old age may at first sight seem surprising.
Indeed,
one can argue that the bond due to a marriage that has lasted
many years should be stronger than the link
between young people who got married just a few years ago.  
If the bond is stronger, then its severance should have a
more dramatic effect, a prediction that comes in contradiction
with the evidence shown in Fig. 3a.
However, to say that the strength of a bond increases in
the course of time is just wishful thinking. In fact,
the attraction between men and women becomes weaker
for old ages; this is
not wishful thinking for it is attested
by the observation that the marriage rate of single people 
(for both males and females) decreases%
steadily after reaching a maximum around the age of 25
\qfoot{The rate is of course computed with respect to the
population of bachelors present at each age; it
shrinks quickly after the age of 25.}%
. 
Hence, if the attraction is stronger for young people it
is natural that the severance of the bond should be more
detrimental.
\qpar

Perhaps some readers may think that attributing the lower
death rate of married people to the existence of a bond 
between them is
a conclusion that is drawn too quickly. After all, there could
be other explanations. For instance one could argue that
the economic situation of married people is on average
better than the situation of non-married people.
In order to answer this kind of objections we provide
two other pieces of evidence.

\qA{Bonds between parents and children}

It seems plausible to expect 
the bond between parents and their
children to be almost as strong as the bond between
husbands and wives. This conjecture is confirmed
by the data in Table 1.

\begin{table}[htb]

 \centerline{\bf Table 1: Effect on suicide rates of
husband-wife bonds and parents-children bonds}

\vskip 3mm
\hrule
\vskip 0.7mm
\hrule
\vskip 2mm

\color{black} 

$$ \matrix{
\qtb
\hbox{Situation} \hfill  & \hbox{M} & \hbox{F} &\hbox{ }&\hbox{M} & \hbox{F}\cr
\noalign{\hrule}
\qth 
\hbox{Married with children} \hfill& 20 & 4.5& & 1 & 1\cr
\hbox{Married without children} \hfill& 47 & 16 & & 2.4&3.6 \cr
\hbox{Widowed with children} \hfill& 52 & 10 & & 2.6& 2.2\cr
\qtb
\hbox{Widowed without children}\hfill & 100 & 23 & & 5.0 & 5.1\cr
\noalign{\hrule}
} $$
\vskip 0.5mm
Notes: The table gives suicide rates (per 100,000 people)
in France in the 8 years between 1861 and 1868. ``M'' means
male, ``F'' means female.
The two columns on the right-hand
size repeat the same data with a different normalization.
If one accepts the hypothesis (made by Emile Durkheim and other
sociologists of the late 19th century) that it is the severance
of bonds and especially of family bonds which is the main social
factor in the phenomenon of suicide, then these data allow
us to compare the respective strengths of the bonds 
between husband and wife on the one hand 
and between parents and children on the other hand.
The fact that the increase in the suicide rate is almost the same
through widowhood or through lack of children suggests that these
bonds have almost same strength.
\qL
{\it Source: Bertillon (1879, p. 474)}
\vskip 3mm
\hrule
\vskip 0.7mm
\hrule

\end{table}
%%-----------------------------------------------

\qA{Death rate by marital status and by cause of death}

The graphs  in Fig. 3b give detailed data by marital status 
and by age for 6 different causes of deaths. The age-group
correspond to 10-year intervals. Needless to say, such 
statistics make sense only for a country with a large population,
otherwise there would be too few cases in each cell
of the table. Just as an illustration, for widowed men
in the 25-34 year age group there were only 54 suicides
and 7 deaths due to cancer.

%%
%%%% MORTALITE PAR ETAT MATRIMONIAL: USA (VITAL STATISTICS)
\begin{figure}[htb]
\centerline{\psfig{width=10cm,figure=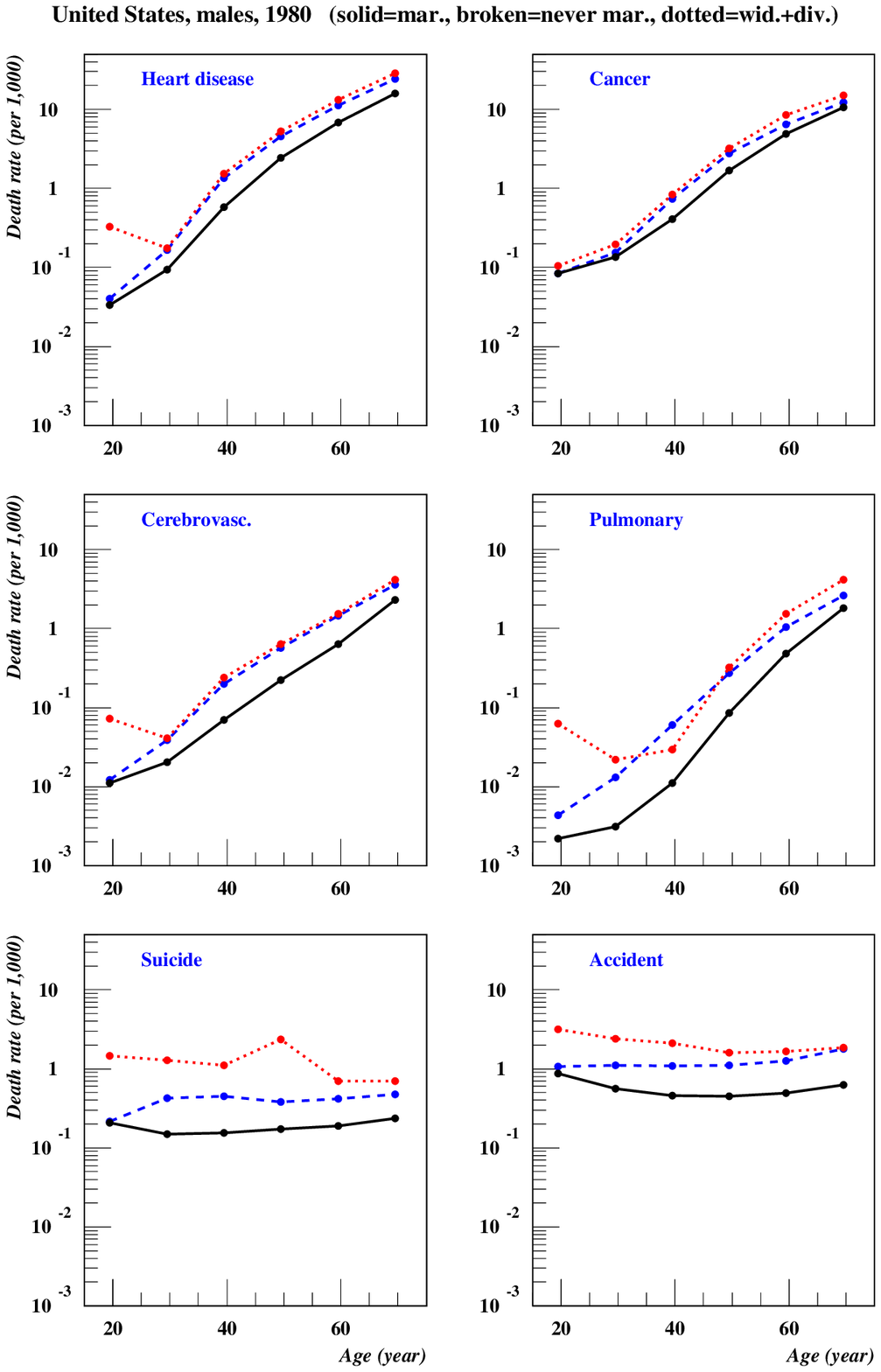}}
\qleg{Fig. \qhu 3b\qhv Mortality rates by cause of death
according to marital status.}
{On average the death rate of non-married persons is about
twice the rate of married persons. For cancer the
ratio is somewhat smaller
but for the suicide of widowers and divorced males it is a factor
of about 5.
There is a similar but somewhat weaker effect for females.
Similar data are available also for 1979 and show very much the same
effect.In terms of life expectancy the higher mortality rate
results in a reduction of about 4-5 years. This fairly small 
reduction is due to the fact that doubling a death rate
that is very low (as is the case until the age of 60-65)
does not produce much difference in life expectancy.}
{Source: Vital Statistics of the United States, 1980.}
\end{figure}
%-------------------------------------------------
%
Quite surprisingly, the excess mortality 
due to the marital status is broadly the same
for various causes of mortality ranging from
heart attack to suicide or accident. So, once again, reduced
ties result in higher mortality. As a matter of fact, these
data raise more questions than they provide answers.
Unfortunately, as far as statistical data are concerned,
one can hardly expect more detailed statistics. Hence,
if one wishes to get a better understanding, one needs
an alternative approach that would allow us to
perform more focused experiments.

\qA{Alternative approach}

That is how the present collaboration
between entomologists and physicists emerged. The active
phase of this investigation started in September 2010. 
Within the past two years a broad but well focused set of experiments
were performed. Each of them lasted about one month and 
many were repeated several times. Our major 
objective was to establish with high reliability 
whether or not the effect really exists. As will be seen
the answer is ``yes''. Moreover, the experiments show that
the effect is not limited to social insects but occurs also
for such non-social insects such as fruit flies%
\qfoot{For the {\it Bactrocera dorsalis} species
that we tested, ``tephritid fruit flies'' would
be the more correct term but, for the sake of simplicity,
this precision will be omitted thereafter.}%
.
\qpar

We realize very well that the present results
are only the  beginning of this interesting story and
that more experiments are needed in order to
identify basic mechanisms. Our hope in writing this
paper is that it can encourage further investigations
in this direction.

\qA{Objectives of the present paper}

In this paper we will try to answer the three following questions.
\qbu It is customary to distinguish 3 types of survivorship
curves. Such a classification is based on curves that give
the logarithm of the proportion of surviving elements as a
function of age. Type 1 curves are flat for most of the life-time
and fall off rapidly in old age.  Human survivorship curves are
of this type. Type 2 curves  fall off linearly; examples will
be given below. Type 3 curves drop sharply in early life
and then level off in later life. To what type belong the
survivorship curves of the ants investigated in this paper?
\qbu Do our experimental results show evidence of a group
effect? More specifically, are the survival rates of 10-ant groups
higher than those of single ants?
\qbu If the answer to the previous question is ``yes'',
does observation suggest a mechanism for this effect.
At this point it would be too early to propose a detailed
mechanism. Instead, we will consider more closely
two types of effects: (i) a ``shock effect'' 
that occurs shortly after the ants are separated 
from their colony.
(ii) A long-term effect that slowly erodes the 
survival likelihood of the ants. While the existence of
the second effect seems fairly expected and unsurprising,
the role of the first one is less obvious. Yet, as will
be seen, it plays a significant role.

%
%%%% MODELE 
%\begin{figure}[htb]
%\centerline{\psfig{width=10cm,figure=FIG/.eps}}
%\qleg{}{}{}
%\end{figure}
%-------------------------------------------------
%

%%%%%%%%%%%%%%%%%%%%
% IMPORTANT! PAS D'ESPACE ENTRE LE \end{minipage} ET LE
%                    \begin{minipage}
\begin{figure}[htb]
\vskip 5mm
% Left-hand side column
\begin{minipage}[t]{0.4\textwidth}
\centerline{\psfig{width=5.5cm,figure=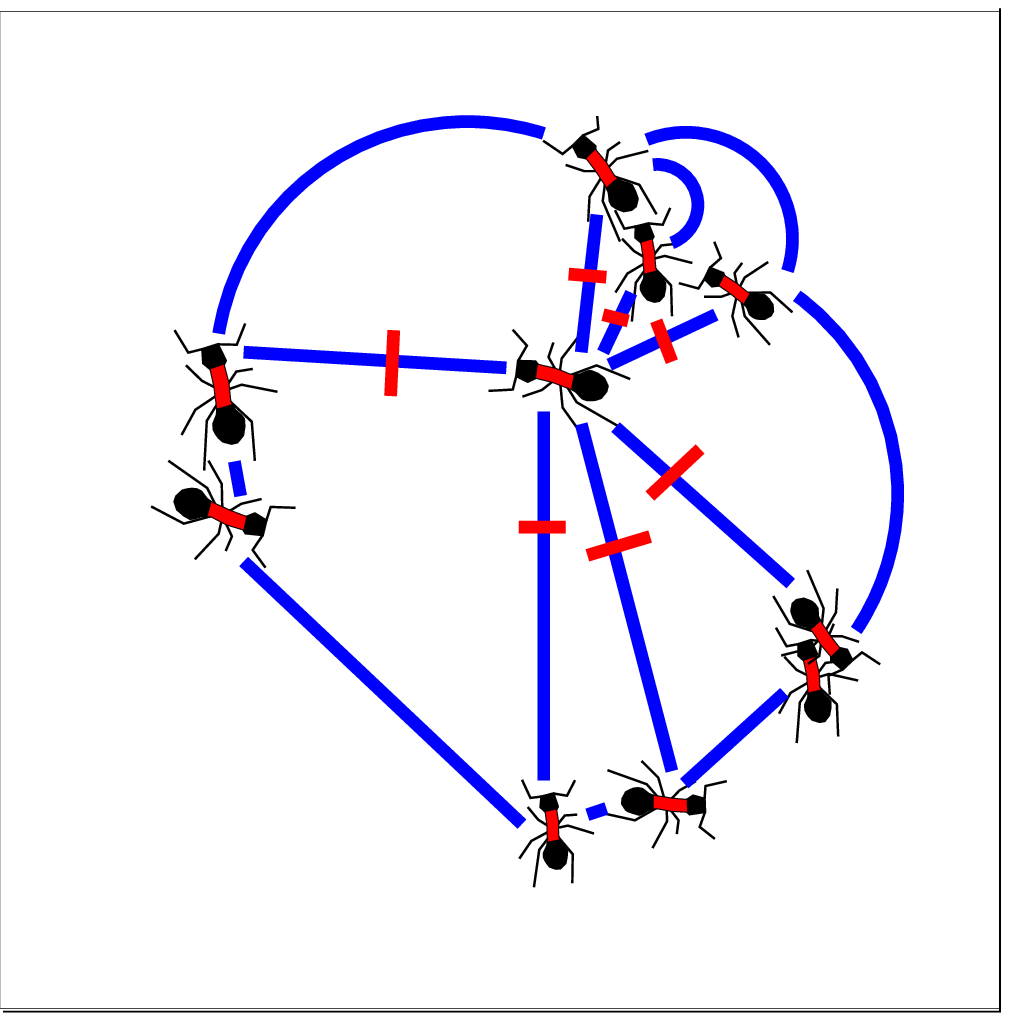}}
\end{minipage}
% Right-hand side column
\begin{minipage}[t]{0.4\textwidth}
\null
\vskip -55mm
\centerline{\psfig{width=6.5cm,figure=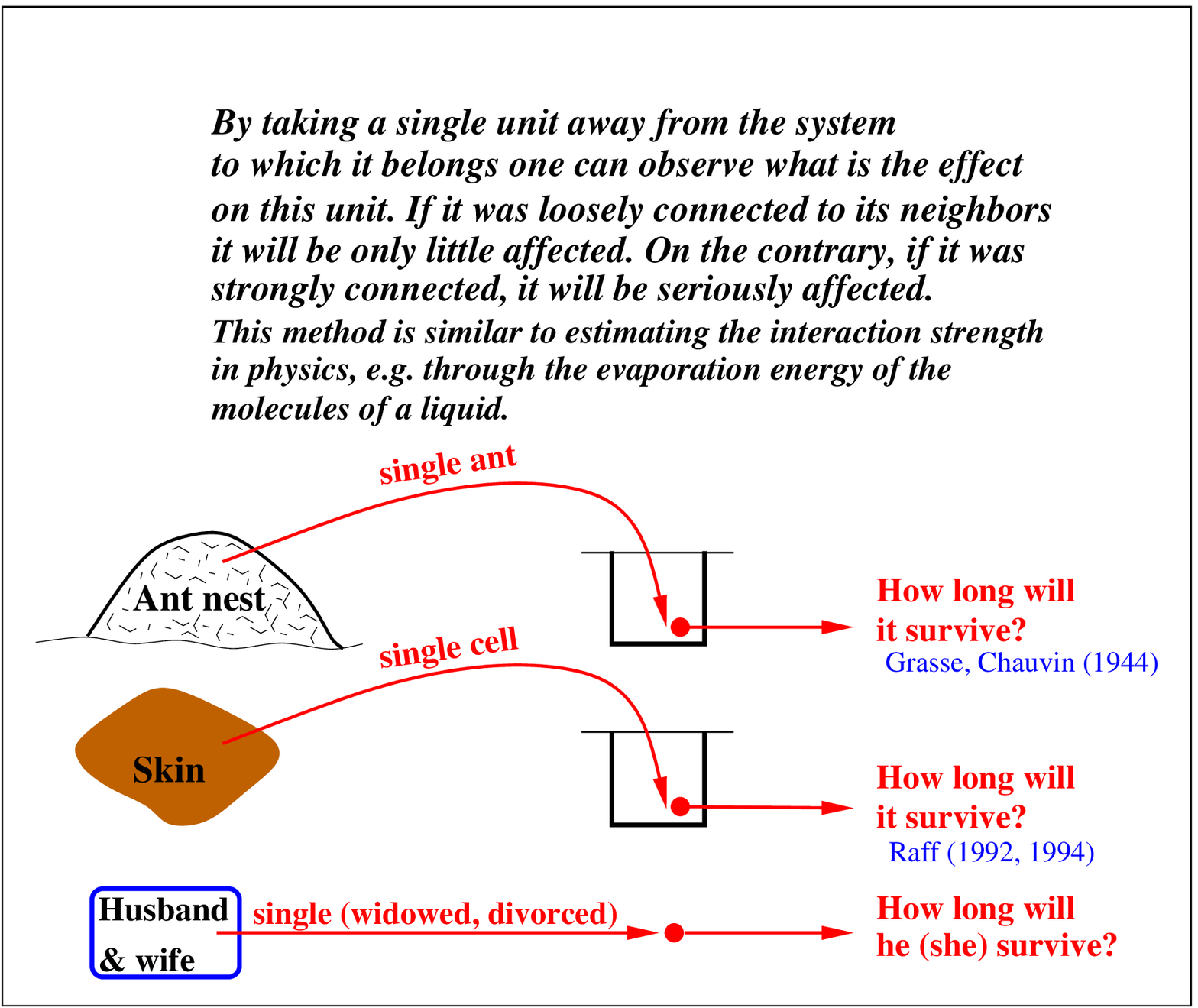}}
\vskip -5mm
\vskip 5mm
\end{minipage}
\vskip 2mm
{\color{blue} \small Fig. 4a (left):  Schematic representation
of the links within a group of ants. 
If (within a range of a few centimeters) the 
interaction between pairs is
not distance dependent, then the interaction within
a group of size $ n $ will increase $ $ approximately
as the number of ties, that is to say as 
$ \pmatrix{n\cr 2\cr}\sim n^2/2 $.
Fig. 4b (right) summarizes the severing of 
the ties that link an unit to its neighbors in different situations.}
\end{figure}
%%%%%%%%%%%%%%%%%%%%%

\qI{Experiments}

In the first part we describe the experiments with 
various species of ants. The experiments with fruit flies are
described in the second part. Then we discuss our procedure in the
light of other possible methods.

\qA{Ants}
The ants used in the experiments were taken from
artificial colonies kept in the laboratory. They were placed
in plastic bowls which had a lower diameter of 4.3cm and
an upper diameter of 10.5cm. To prevent the ants from escaping
the upper sides of the bowls were painted with Fluon%
\qfoot{It is important to understand how Fluon works for
otherwise one might be tempted to think that it is through
the toxicity of its chemicals that it repulses ants.
Needless to say, such a toxicity could affect the experiments.\qL
Fluon is a white liquid containing small (less than
one micrometer in diameter) particles of a solid substance called
polytetrafluorethylene,
a non-stick material better known under its brand name of Teflon.
It has been used to control the movements of  ants as early as 
1956 (see Merton 1956). It works by preventing ants from
using the adhesive pads that they have at the end of their legs.
The role of this pads becomes clear by observing that ants which
have been made to walk on French chalk can no longer climb up
a vertical glass tube (Merton 1956, p. 219). In short,
toxicity does not seem to play any role.}%
.
\qpar

With only one ant per bowl the density will be
7 ants per square decimeter on the bottom of the bowl.
With 10 ants per bowl the density will naturally be 10
times higher that is to say 70 ants per square decimeter.
In other words, this means that each ant will have an area of
1.4 square centimeter. When ants form clusters their density
can easily
be of the order of 20 to 30 per square centimeter that is to
say some 35 times higher.
\qpar

The different replicates (30 were used in the first experiment but
in subsequent experiments there were usually about 10) were 
placed on tables with a distance of 60cm between adjacent bowls.
Thus, altogether the 60 bowls occupied a surface of 21 square meters.
Each contained a tampon with sugar water which was changed
every week. 

\qA{Average death rate of ants in their colony}
What is the lifespan of the ants in normal conditions
that is to say in their nest? In the case
of the red fire ants ({\it Solenopsis invicta}) 
one can distinguish three subgroups of workers
according to their size: minor,
medium and major ranging from 1.5mm to 3mm. For each
kind estimates of their average lifespan are 45 days,
75 days and 135 days respectively%
\qfoot{Website of the University of Florida Institute of
Food and Agricultural Sciences.}%
.
In order to get an overall average life span for the whole
population we need to estimate the proportion of each subgroup
in our colonies. To this end several samples containing
between one and two hundred ants were taken, photographed
and counted. This lead to the following percentages (the
coefficient of variation for the various
samples was of the order of 30\%):\qL
\centerline{ minor: 62\%,\quad medium: 28\%,\quad major: 10\%}
which in turn gave the following weighted average for the lifespan:\qL
\centerline{ Average lifespan of red fire ants $ L \simeq $ 62 days}
\qpar
What is the average mortality rate corresponding to this
life expectancy? $ L $ gives a daily death probability of $ 1/L $;
consequently, the death rate per day and per 1,000 ants
will be $ 1000/L= 16 $. We will see later that the average death rate
of isolated ants is about 2 or 3 times higher.

\qA{Fruit flies}

The fruit flies {\it Bactrocera dorsalis} were reared in
the laboratory. As they 
are about 3 to 4 times
bigger than the ants, they needed larger containers.
Moreover, in order to prevent them from flying away the cages
had to be closed. The cages were (30cm, 30cm, 30cm) cubes with wooden
frames and sides consisting of gauze except for
the top side which had Plexiglass to allow easy inspection. 
The fruit flies were adult unmated females. They were given water,
yeast and sugar. There were 30 replicates with a spacing of
60cm between the cages.
\qpar

The bowls and cages were inspected every 12 hours.

\qA{Experimental procedures}
In what will be called the main experiment there
were two kinds of cells: those containing just one worker,
and those with 10 workers.
It must be observed that in addition to this experiment,
different variants were carried out which involved the following
changes.
\qee{1} In variant (1), a queen was added to the {\it Solenopsis
invicta} workers. Thus, there were two kinds of cells: those
containing 1 worker + 1 queen and those with 10 workers + 1 queen.
\qee{2a} In variant (2a) the ants received water but no sugar.
\qee{2b} In variant (2b) the ants received neither water nor sugar.
\qpar

The main purpose of experiments (2a) and (2b) was to
make the experiment faster. Whereas from the start to 
the death of all ants the main experiment lasted
almost 30 days%
\qfoot{For {\it Bactrocera} the duration was even much longer.
After 40 days, 30\% of the single flies and 12\% of those in groups
of 10 had died. Then, between day 40 and day 70 for some
unknown reason there was a rapid increase of the mortality
in the 10-fly groups. On day 70
when the experiment was stopped one half of the
single flies and one fourth of flies in the 10-fly groups were still
alive.}%
,
variant (2a) lasted only 8 days for {\it Solenopsis} and 2.5 days for
{\it Bactrocera}. Not surprisingly, variant (2b) was even shorter:
2 days for {\it Solenopsis} and 2 days for
{\it Bactrocera}. These experiments were inspired by those
done in December 1972 - January 1973
by R\'emy Chauvin on bees (Chauvin 1973). Chauvin observed
that for single bees death rates were some 2 to 3 times higher 
than in groups of 10, a result which was consistent with those
obtained previously by Grass\'e and Chauvin in 1944 (Fig. 1).
This supported the hope that meaningful results could be
obtained in a much shorter time than in the main experiment. 
\qpar

However, the experiments that we have done for
{\it Solenopsis} and {\it Bactrocera}
did not lead to any clear results. Altogether 5 graphs were 
drawn: \qL
{\it Solenopsis}: (water, no sugar) + (no water, no sugar)\qL
{\it Bactrocera}: (water, no food) + (no water, food) + (no water, no
food)\qL
In all cases except one the death rates of the two sorts of groups
were almost the same. The only case which showed a difference
was {\it Solenopsis}, (water, no sugar). In this case,
after 8 days the cumulative deaths in the 10-ant groups was
2.3 times {\it higher} than the deaths in the 1-ant group.
These mostly meaningless results lead us to drop this
methodology%
\qfoot{It can be added that the necessity of
making the insects starve was fairly unpleasant which explains
that we were rather glad to forget this method.}%
.
\qpar

Whereas the 1-ant cells were just removed
after the death of the ant, for the 10-ant cells
several experimental procedures were possible.
\qbu Each dead insect could be replaced by a new living 
insect so as to
maintain the same number of 10 throughout the experiment. 
\qbu The dead ants could be left in the bowls. 
\qbu The dead ants could be removed at time of inspection
which means that their corpses may remain in the container
for a maximum of 12 hours.
\qpar
It is the last option which was selected. 
Of course, it would
be interesting to see whether or not the two other options
lead to similar results. This is a point which may be investigated
in the future.

\qpar
The main features of the different experiments are summarized in
table 1.

%%-----------------------------------------------
\begin{table}[htb]

 \centerline{\bf Table 2: Summary of the experiments}
\small

\vskip 3mm
\hrule
\vskip 0.7mm
\hrule
\vskip 2mm

\color{black} 

$$ \matrix{
&\hbox{Date} \hfill  & \hbox{Insect} & \hbox{Species} \hfill & \hbox{Average} &
\hbox{Replic.}& \hbox{Replic.}&\hbox{Mortality} \cr
&\hbox{} \hfill  & \hbox{} & \hbox{} & \hbox{size} &
 & &\hbox{ratio} \cr
\qtb
&\hbox{} \hfill  & \hbox{} & \hbox{} & \hbox{[mm]} &
\hbox{}& & \hbox{} \cr
\noalign{\hrule}
\qtb
&\hbox{} \hfill  & \hbox{} & \hbox{} &  &
  &  &  \cr
&\hbox{} \hfill  & \hbox{} & \hbox{} & \hbox{} &
\hbox{1 W}& \hbox{10 W}&  \hbox{(1)/(10)}\cr
\hfill 1&\hbox{25 Sep 2010} \hfill  & \hbox{ant} & \hbox{Solenopsis inv.}  \hfill& 2 &
 30 & 30 & 2.25 \cr
\hfill 2&\hbox{18 Dec 2010} \hfill  & \hbox{ant} & \hbox{Solenopsis inv.}  \hfill& 2 &
 9 & 9 & 1.10 \cr
\hfill 3&\hbox{12 Jan 2011} \hfill  & \hbox{ant} & \hbox{Solenopsis inv.}  \hfill& 2 &
 10  & 10 &  0.93\cr
\hfill 4&\hbox{6 Sep 2012} \hfill  & \hbox{ant} & \hbox{Solenopsis inv.}  \hfill& 2 &
 10 & 10 & 1.95 \cr
&\hbox{\bf \color{blue} Average (1-4)} \hfill  & \hbox{} & \hbox{} & \hbox{} &
& & \hbox{\bf \color{blue} 1.80}{\color{blue} \pm 0.3}\cr
&\hbox{} \hfill  & \hbox{} & \hbox{} & \hbox{} &
& & \hbox{} \cr
&\hbox{} \hfill  & \hbox{} & \hbox{} & \hbox{} &
\hbox{1 W}&\hbox{10 W} & \hbox{(1)/(10)} \cr
\hfill 5&\hbox{18 Dec 2010} \hfill  & \hbox{ant} & \hbox{Pheidole pall.}  \hfill& 2 &
  13 &  10 &  2.12\cr
\hfill 6&\hbox{18 Dec 2010} \hfill  & \hbox{ant} & \hbox{Tapinoma mela.}  \hfill& 1.5 &
  23 &  11 &  0.70\cr
&\hbox{\bf \color{blue} Average (1-6)} \hfill  & \hbox{} & \hbox{} & \hbox{} &
& & \hbox{\bf \color{blue} 1.62}{\color{blue} \pm 0.2} \cr
&\hbox{} \hfill  & \hbox{} & \hbox{} & \hbox{} &
\hbox{}& & \hbox{} \cr
&\hbox{} \hfill  & \hbox{} & \hbox{} & \hbox{} &
\hbox{1 fly}& \hbox{10 flies}& \hbox{(1)/(10)} \cr
\hfill 7&\hbox{15 Nov 2010} \hfill  & \hbox{fruit fly} & \hbox{Bactrocera dors.} \hfill & 8 &
  20 &  8 &  \hbox{\bf \color{blue} 2.12}\cr
&\hbox{} \hfill  & \hbox{} & \hbox{} & \hbox{} &
\hbox{}& & \hbox{} \cr
&\hbox{} \hfill  & \hbox{} & \hbox{} & \hbox{} &
\hbox{1W+Q}& \hbox{1W}& \hbox{(1W,Q)/(1W)} \cr
\hfill 8&\hbox{18 Dec 2010} \hfill  & \hbox{ant} & \hbox{Solenopsis inv.}  \hfill& 2 &
 9 &  9 &  2.00\cr
\hfill 9&\hbox{12 Jan 2011} \hfill  & \hbox{ant} & \hbox{Solenopsis inv.}  \hfill& 2 &
 10 &  10 &  1.43\cr
\hfill 10&\hbox{6 Sep 2012} \hfill  & \hbox{ant} & \hbox{Solenopsis inv.}  \hfill& 2 &
 10 &  10 &  1.00\cr
&\hbox{\bf \color{blue} Average (8-10)} \hfill  & \hbox{} & \hbox{} & \hbox{} &
& & \hbox{\bf \color{blue} 1.48}{\color{blue} \pm 0.3} \cr
&\hbox{} \hfill  & \hbox{} & \hbox{} & \hbox{} &
\hbox{}& & \hbox{} \cr
&\hbox{} \hfill  & \hbox{} & \hbox{} & \hbox{} &
\hbox{10W+Q}& \hbox{10W} & \hbox{(10W,Q)/(10W)} \cr
\hfill 11&\hbox{18 Dec 2010} \hfill  & \hbox{ant} & \hbox{Solenopsis inv.}  \hfill& 2 &
 9 &  9 &  6.00\cr
\hfill 12&\hbox{12 Jan 2011} \hfill  & \hbox{ant} & \hbox{Solenopsis inv.}  \hfill& 2 &
 10 &  10 &  1.30\cr
\hfill 13&\hbox{6 Sep 2012} \hfill  & \hbox{ant} & \hbox{Solenopsis inv.}  \hfill& 2 &
 10 &  10 &  2.00\cr
\qtb
&\hbox{\bf \color{blue} Average (11-13)} \hfill  & \hbox{} & \hbox{} & \hbox{} &
& & \hbox{\bf \color{blue} 3.10}{\color{blue} \pm 1.5} \cr
\noalign{\hrule}
} $$
\vskip 0.5mm
Notes: All mortality ratios were computed 20 days after the
experiments started even when the experiment lasted longer; this
ensured that the number of ants in the groups of 10 did not
fall off too much. The error bar indicated after $ \pm $ is the standard
deviation (which corresponds to a confidence level of 0.68).
Each experiment lasted about one month.
The dates refer to the end of the experiments. The replication columns
refer to the number of groups of each kind that were observed in 
parallel. W means worker, Q means queen.
The size of the insects was indicated because it is well known that
bigger insects survive longer
but one does not expect size to play a role in the mortality ratio.
\qL
{\it Source: Experiments done at South China Agricultural University.}
\vskip 3mm
\hrule
\vskip 0.7mm
\hrule

\normalsize

\end{table}
%%-----------------------------------------------

\qI{Experimental results for groups with 1 versus 10 workers}

The figures 5 and 6 give detailed results for each separate
experiment. Why did we choose to give separate results
rather than just global averages? The reason is very simple.
For most experiments the groups of 10 had a lower mortality than
the singles, yet there were a few exceptions. We do not
wish to hide such exceptions behind global averages (which will
be given later). Needless to say, the fact that a replication
gives a result which contradicts the initial experiment is
quite intriguing and one would like to know the reason for that.
This is not easy, however, because there are many possible
explanations and in order to determine which one is
correct one needs to perform numerous control experiments.
\qpar
Just to give an idea of the kind of effects that may be involved,
we recall a similar problem which arose in survival experiments
done by R\'emy Chauvin (1952). In 1952 he tried to replicate
his experiment of 1944 (summarized in Fig. 1).
In this second experiment the bees
received exclusively sugar water with 20\% sugar. Chauvin
was very surprised to see that the results were opposite to what he
had observed in 1944 in the sense that the mortality was higher in
the groups with many individuals than in those with only a few%
\qfoot{Actually, Chauvin should not have been so surprised.
Indeed, what is very clear from Fig. 1 is that the bees die
{\it quickly}; however, the difference between groups of 1,2 or 10 
is fairly small. In other words, the shock of being cut off from
their colony is so strong that even a few neighbors make
no difference.}%
.
\qpar
Eventually, Chauvin came up with the following explanation.
The bees in large groups were more active and therefore absorbed
more sugar water. However, for some reason, it seems that the 
sugar syrup made them ill. Their abdomen was inflated and eventually
they died. It is possible that some kind of bacteria had
developed in the syrup. It was changed only every 5 days. 
So the more active the bees, the more of the deadly syrup
they absorbed and the quicker they died. 
\qpar
This story also shows the limit of control experiments.
If no bacteria develop during the control experiment 
the result will be different once again. 

%%%%%%   DEUX GRAPHES AVEC BARRES D'ERREUR
\begin{figure}[htb]
\centerline{\psfig{width=14cm,figure=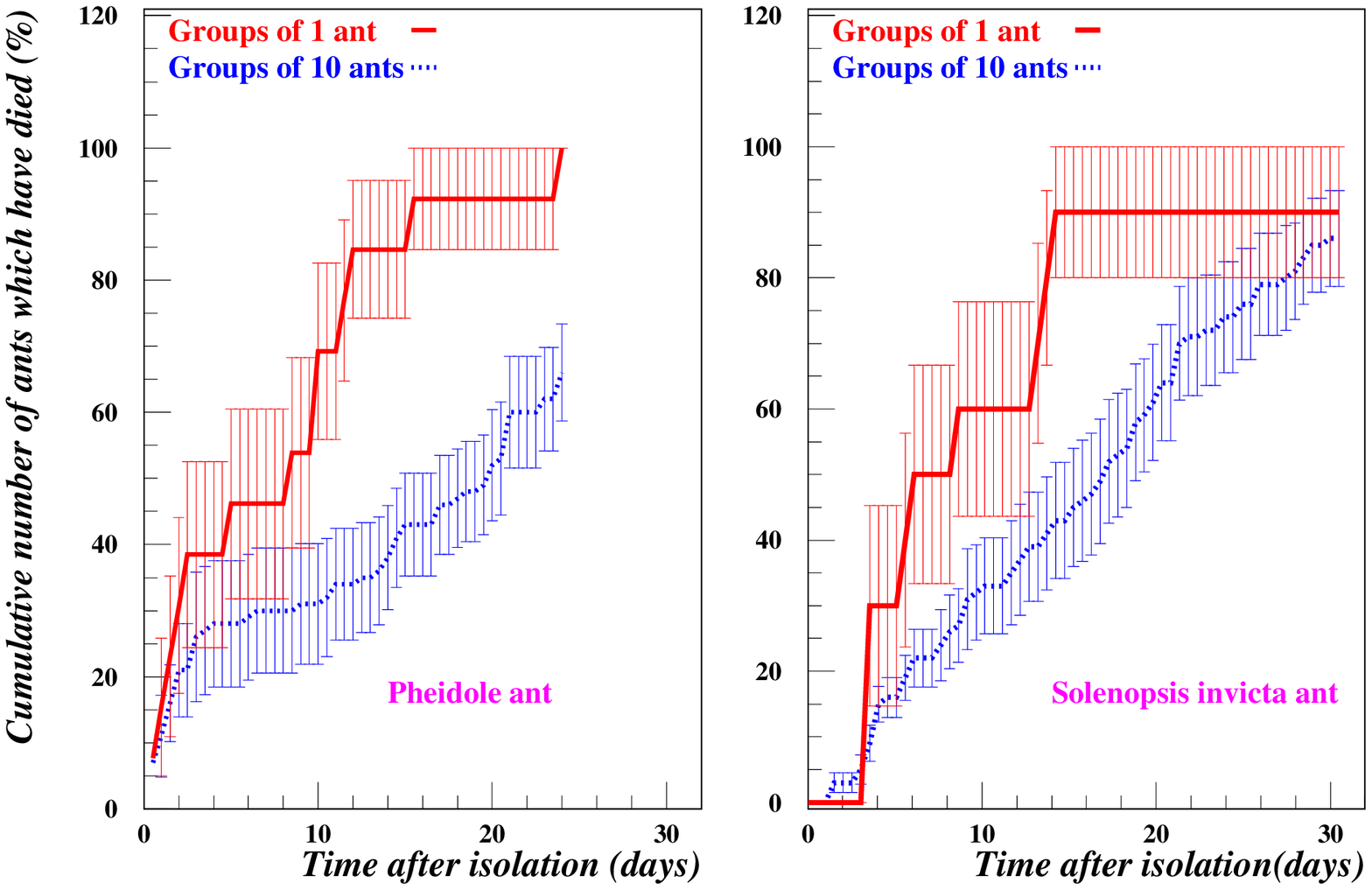}}
\qleg{Fig.\qhu 5\qhv Two examples of graphs with error bars.}
{For the left-hand side experiment (which corresponds to number 5
of Table 2) there were 13 groups of 1 ant and 10 groups of 10 ants.
For the experiment on the right-hand side (which corresponds
to number 4 of Table 2) there were
10 groups of 1 ant and 10 groups of 10 ants.
At each time step the standard deviation $ \sigma $
of the cumulative 
number of deaths in each box was computed. The error bars
show $ \pm \sigma $. Under the (standard) assumption of a Gaussian
distribution this corresponds to a probability confidence level
of 0.68. It is customary to say that one curve (the red one) is 
{\it significantly} higher than another (the blue one)
when their error bars do not overlap.
Of course, such a statement is confidence level dependent.
For a confidence level of 0.99 the error bars would be
2.58 times wider which would make them overlap.
Thus, it can be said that our conclusion holds with probability
0.68 but to confirm it with probability 0.99 would necessitate
additional observations.}
{Source: The experiments were performed at the South China
 Agricultural University in November 2010 and September 2012
respectively.}
\end{figure}
%-------------------------------------------------
%

%%
%%%%%%   LES 6 GRAPHES POUR GROUPES DE 1 OU 10
\begin{figure}[htb]
\centerline{\psfig{width=14cm,figure=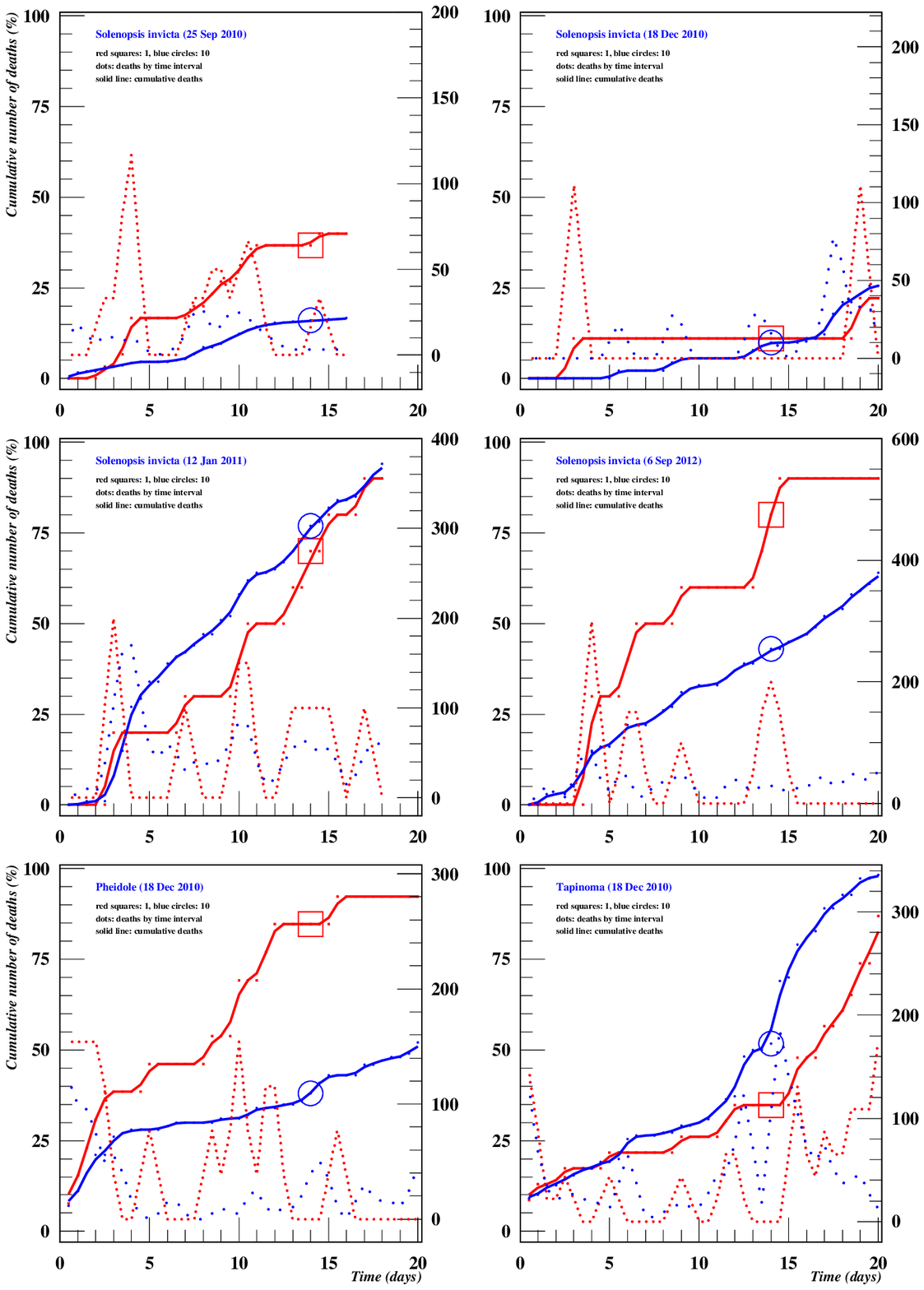}}
\qleg{Fig.\qhu 6\qhv Survival experiments with 1 or 10
workers per group.}
{The solid line curves give the cumulative numbers of
deaths whereas
the dotted curves give the death rates per 1,000 individuals
in successive 24-hour intervals
that is to say the
number of deaths during such intervals divided by
the living population at the beginning of the interval.
Graphs 1,2,3 and 4 are replications of the fire ant
experiment, while graphs 5 and 6 concern two different
species of ants.
It can be observed that there is often a big surge of
deaths at the beginning of the experiment especially for
single ant groups.}
{Source: These experiments were done at the South China
Agricultural University at various times between September 2010
and September 2012.}
\end{figure}
%-------------------------------------------------

%%
%%%%%%  BACTROCERA DORSALIS (FRUIT FLY)
\begin{figure}[htb]
\centerline{\psfig{width=16cm,figure=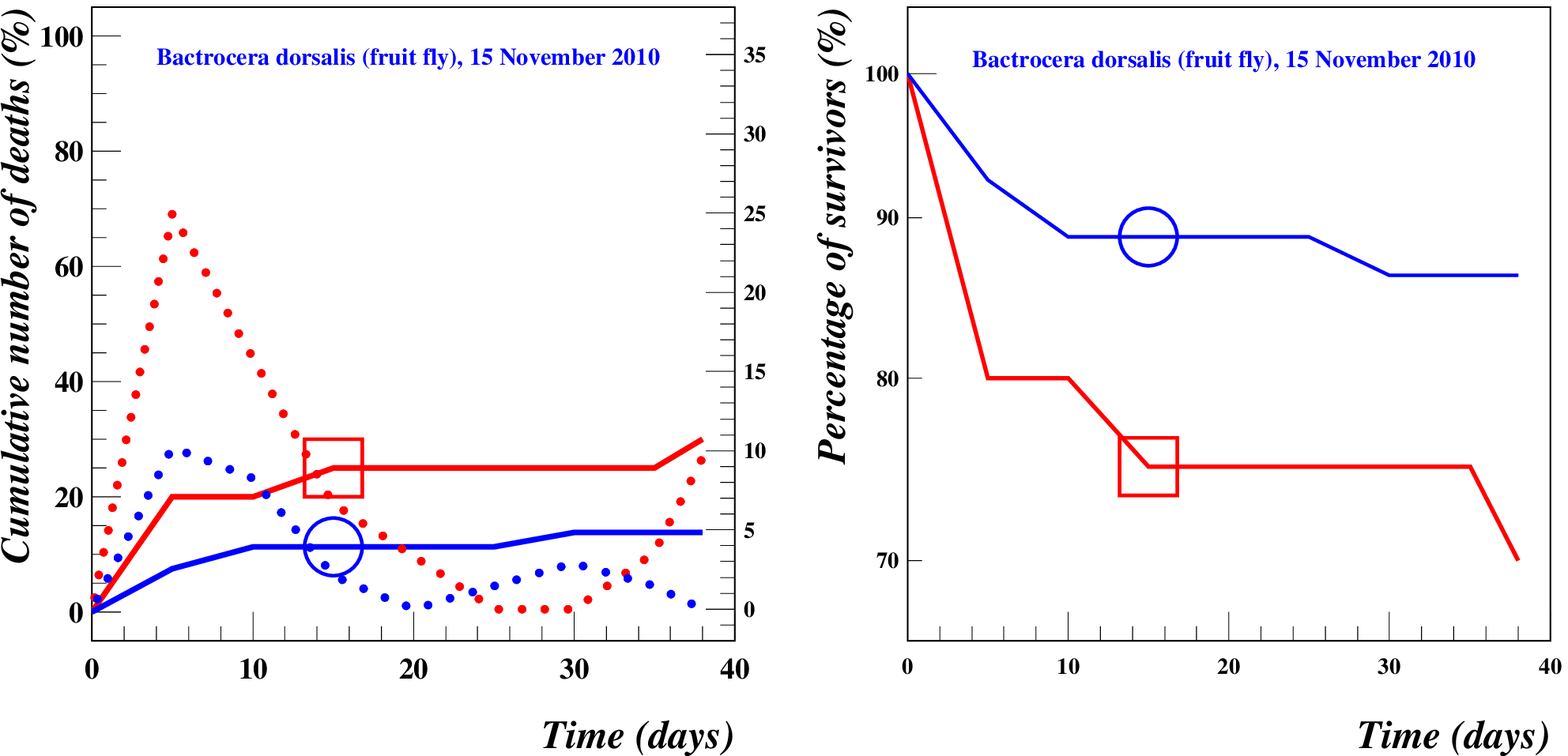}}
\qleg{Fig.\qhu 7\qhv Survival experiment for fruit flies.}
{As in the previous graphs, the red solid line with the square
represents single-individuals (20 groups)
while the blue solid line with the circle represents the 
10-flies groups (8 groups). The dotted curves represent
the daily death rates per 1,000 individuals.
Compared with the ant experiments,
the mortality is much lower. 
The dotted-line curves show the
daily mortality rate (expressed in number of deaths per insect).
It can be seen that there is a big mortality burst at the 
beginning of the experiment. After that the mortality 
is very low. \qL
The graph on the right-hand side shows the survivorship curve.
It belongs to type 3. In fact, the experiment was
continued until day 70. After day 45, for some unknown
reason, the mortality in 
the groups of 10 started to grow and eventually their cumulative
number of deaths crossed and surpassed the curve of singles.}
{Source: This experiment was done at the South China
Agricultural University in October-December 2010.}
\end{figure}
%-------------------------------------------------

\qI{Experimental results for groups with/without queen}

The experiments considered in this section are similar to
the fire ant experiments 
done in the previous section except that a queen was
added to each of the groups. What was our purpose
in adding queens?
\qpar

First of all, it must be recalled that fire ant colonies
have several queens. For species whose colonies contain only
one queen it is a common idea that the queen is a crucial
node in the global colony network (although we do not know
precisely how this node is integrated in the rest of
the network). Even for species whose colonies have several
queens it is natural to assume that in the colony's organization
the queens play a more important role than ordinary workers.
In adding a queen to the isolated groups one may expect
them to become mini-colonies in their own right.
This idea was particularly plausible for groups of 10 ants.
which is why our analysis will mainly focus on this case.
Under the assumption that groups of 10 ants plus one queen
become mini-colonies one expect them to survive longer
than groups of 10 without queens. 
\qpar
However, as seen in the graphs that are on the right-hand side
of Fig. 8, the experiments showed
exactly the opposite. That was a big
surprise. How can one interpret this result?
\qpar

%%
%%%%%%   LES 6 GRAPHES AVEC ET SANS ``QUEEN''
\begin{figure}[htb]
\centerline{\psfig{width=10cm,figure=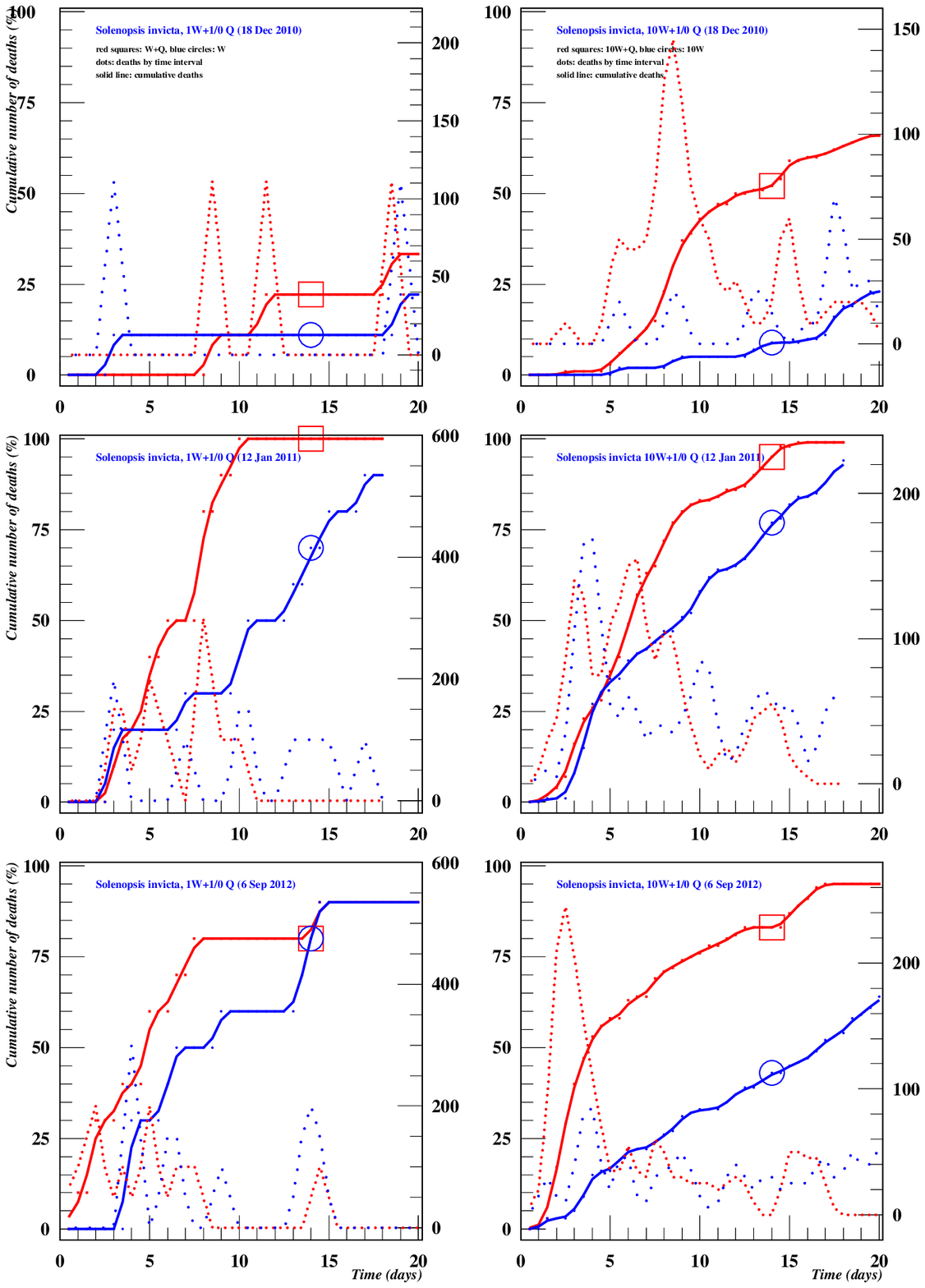}}
\qleg{Fig.\qhu 8\qhv Survival experiments with or without queen
in the groups of workers.}
{The graphs in the first column give a comparison of groups of 1 worker
with or without queen. The second column compares groups of
10 workers with or without queen. It should be noted that only 
the deaths of the workers were recorded, so the proportions refer
to the ratios: (deaths of workers)/(Initial number of workers).
the dotted curves represent daily death rates per 1,000 individuals.}
{Source: The experiments were performed at the South China
Agricultural University in November-December 2010 and in
September 2012.}
\end{figure}
%-------------------------------------------------
%

When an element has very few interactions with its neighbors, 
its isolation will make little difference; hence, one 
expects little or no reduction in life expectancy.
On the contrary, for an element which interacts strongly
with its neighbors isolation will be a great shock; therefore
one would not be surprised to see a sharp reduction in life
expectancy. This is schematized in Fig 4a,b.
The conclusion which derives from this
argument is that the groups of 10 ants plus one queen 
have a stronger interaction with the rest of the colony than
the groups of 10 ants without queen. This is a natural
conclusion in spite of the fact that, so far, we do not 
understand the mechanism of this interaction.

\qI{Systems science perspective about survivorship}

In the following sections we will draw parallels between systems 
which, at first sight, seem to be
very different. We realize very well that in current conceptions
such comparisons 
may appear questionable and far-fetched. However, we believe
that such reservations more due to our own peculiar position
in the observation process 
rather than to objective factors. 
The key-point is that systems of living organisms
are of macroscopic dimension which means that we
can easily observe them. In physics we easily accept that the same
rules applies to very different objects because we do
not realize how dramatically different they are. This results
from our inability to observe them directly.
Before we discuss survivorship curves
for ants and other organisms we wish to explain and illustrate this
important point through a specific case.

\qA{Binding energy in systems science perspective}

What makes the strength of physics is the fact that
its laws have a broad range of validity. 
An obvious example is the law of free fall
that we already mentioned at the beginning.
A hazelnut or an apple have little in common with the Moon, yet
their movements are governed by the same law.
\qpar 
Here is another illustration that is not so well known 
but is more germane to the topic of this paper.
There is a basic rule which says that the creation of 
links is an exothermic process whereas breaking existing
links requires heat to be applied to the system and
is therefore an endothermic process.

\qA{Water molecules versus protons and neutrons in nuclei}
A well-known
illustration is the boiling of water in which the
links between water molecules are severed. A less-trivial
example is the mixing of water and ethanol. 
If the mixing is done with same volumes of water and
ethanol it evolves about 200 calories (48 kJ)
per mole of mixture
(Bose 1907). This release of heat is the consequence of the
fact that the
mixing results in strong hydrogen bonds being
established between water and ethanol molecules%
\qfoot{Actually, the heat that is released is the difference 
between the energy that it takes to push the water and ethanol
molecules apart and the energy that is produced by the
creation of the water and ethanol molecules. The fact
that heat is released rather than absorbed shows that the 
latter is larger than the former.}%
:
 $$ \hbox{H}_{\small 2}\hbox{O} + 
\hbox{HO}-\hbox{CH}_{\small 2}\hbox{CH}_{\small 3} \longrightarrow
\hbox{H}_{\small 2}\hbox{O} ---  
\hbox{HO}-\hbox{CH}_{\small 2}\hbox{CH}_{\small 3} + 48\hbox{kJ}
\qn{1} $$

Yet another completely different case where the same
rule applies is the ``mixing'' of lithium and deuterium
nuclei%
\qfoot{Li designates lithium, the second nucleus is deuterium.
The upper figure is the number of protons and neutrons,
the lower fifure is the number of protons. The two nuclei which
are produced are so-called $ \alpha $ particles. More details
can be found in the following Wikipedia article:\qL
http://en.wikipedia.org/wiki/Nuclear\_reaction}%
:
$$ {\hbox{\phantom X}}_3^6\hbox{Li}\ +
{\hbox{\hskip -2mm\phantom X}}_1^2\hbox{H}
\longrightarrow 2{\hbox{\hskip -2mm \phantom X}}_2^4\hbox{H}+2\ 10^9 \hbox{kJ}
\qn{2} $$

Writing these reactions in the same form as in (1) and
(2) conveys the
impression that the two cases are fairly similar. 
In fact, there are huge qualitative differences between them
as we will see now.
\qbu First, we can see that the amount of energy that is released
is tremendously higher for reaction (2). In fact, even for a 
nuclear reaction the energy produced in (2) is particularly high.
This is due to the fact that the
$ \alpha $ particles that are created have a very high binding
energy per nucleon. In this sense (2) provides a good
illustration of the general rule.
\qbu One should realize that 
in (2) the energy is released in completely different ways than in (1). 
Whereas in (1) the energy is released as heat, in (2) 
it is produced
in the form of (i) high energy light rays called gamma rays
(ii) kinetic energy of the $ \alpha $ particles. It is
because we {\it now} know that these effects are just different
forms of energy that we can establish a parallel between (1) and
(2).
\qbu Needless to say, the lithium, deuterium and $ \alpha $ 
nuclei are
very different from the water and ethanol molecules. 
First, there is a huge difference in size.
The diameters of
water and ethanol molecules are 200 picometer (1pm= $ 10^{-12} $m)
and 500pm respectively%
\qfoot{In ethanol, the average distance between two molecules
is about 700pm (by dividing the volume of one mole by Avogadro's
number one get the volume of the space allowed to each molecule
from which the previous value derives easily).}%
,
whereas the diameter of an $ \alpha $ particle is $ 4\ 10^{-3}\ $pm.
In other words, the difference in size is much larger than between
ants and humans. 
\qbu 
In terms of interaction, the nuclei of reaction (2) are
also much different from the molecules of reaction (1). 
The intermolecular attraction is an electrostatic force
that depends on the spatial distribution of electrical charge of
the molecules whereas 
the protons and neutrons of the nuclei are held
together by the so-called nuclear force which is independent 
of electric charge. 
\qpar

Before we close this discussion we must devote a few words to
an apparent paradox that will certainly come to the mind
of the reader. It is the fact that {\it some} fission reactions
can release huge amounts of energy as seen in the 
atomic bomb.

\qA{Explanation of the paradox of fission reactions.}

At first sight, the rule stated above
seems to be contradicted
by nuclear fission reactions. As one knows, by breaking up
a heavy nucleus such as uranium-235 or plutonium-239
into two smaller nuclei one can obtain the release
of a substantial amount
of energy. Such energy is used in nuclear bombs and reactors.
\qpar

However, this effect is due to what can be called an ``anomaly''.
It results from a competition between two forces: on
the one hand there is
the nuclear force which binds together the protons and neutrons
but has
a very short range and on the other hand there is the 
electrostatic force which is a repelling force (at least 
for the protons) of much larger
range and which therefore tends to dominate for large
nuclei. This explains why most of the large nuclei are in fact
unstable. 
\qpar

This explanation can equivalently be presented
in terms of binding energy. By convention a bound state
such as a nucleus has a {\it negative} energy%
\qfoot{This is because of the convention which attributes
a zero potential energy to the state in which
all constituents are far apart from one another.}%
whose absolute value
is the binding energy, i.e. the energy that must be applied
to separate all components. The higher the binding energy
per nucleon the more stable is the nucleus. 
As a function of nucleus size this maximum
occurs in the vicinity of iron. Most nuclei which are bigger than
iron nuclei are unstable and that instability increases as
the nuclei become bigger. It is because of this anomaly that
a fission reaction can release energy.
\qpar
In contrast, the phenomenon of fusion of two light nuclei
is a more ``normal'' nuclear process because in this case
the electrostatic force is negligible.
\qpar

In the early 20th century when physicists started to study nuclear
reactions they were altogether ignorant about nuclei and nuclear
forces and almost as ignorant about intermolecular forces. 
In a sense that is what allowed them to make bold assumptions.
Nobody was shocked because nobody had a clear understanding
of the qualitative differences between molecules and nuclei. 
On the contrary, the organization of ant colonies has been
known for at least three centuries (e.g. see the work of 
the French naturalist R\'eaumur%
\qfoot{René-Antoine Ferchault de Réaumur (1683-1757)
wrote a ``Histoire des Fourmis'' [Natural history of ants]
which was published only in 1928.}%
).
Needless to say, our knowledge of human societies is even
much older. However, knowledge does not necessarily imply
understanding. Our point is that too detailed knowledge
may be an obstacle to overall understanding.

\qI{Survivorship curves}

Mortality curves are fairly erratic. On the contrary,
survivorship curves which are derived from the former by 
the operation of running summation (which for a continuous
variable would correspond to an integral),
are much smoother. That is why three standard types could be
defined which are shown in Fig. 1.
\qpar
In this section, we discuss the survivorship curves
observed in our experiments. For comparative purpose
we also consider survivorship
curves for a number of other cases in the hope of 
throwing some light on our observations.

\qA{Survivorship curves for ants are of type II}

At the beginning of the paper we mentioned that it is
customary to distinguish three types of survivorship curves
but we did not yet provide examples nor did we say to
which type belong the survival charts given in the
previous section. Fig. 9 shows that all ant survivorship
curves recorded in our experiments are of type II.
On the contrary the curve for {\it Bactrocera dorsalis}
is of type III.

%%
%%%%%%   SURVIVORSHIP CURVES (MOYENNES)
\begin{figure}[htb]
\centerline{\psfig{width=15cm,figure=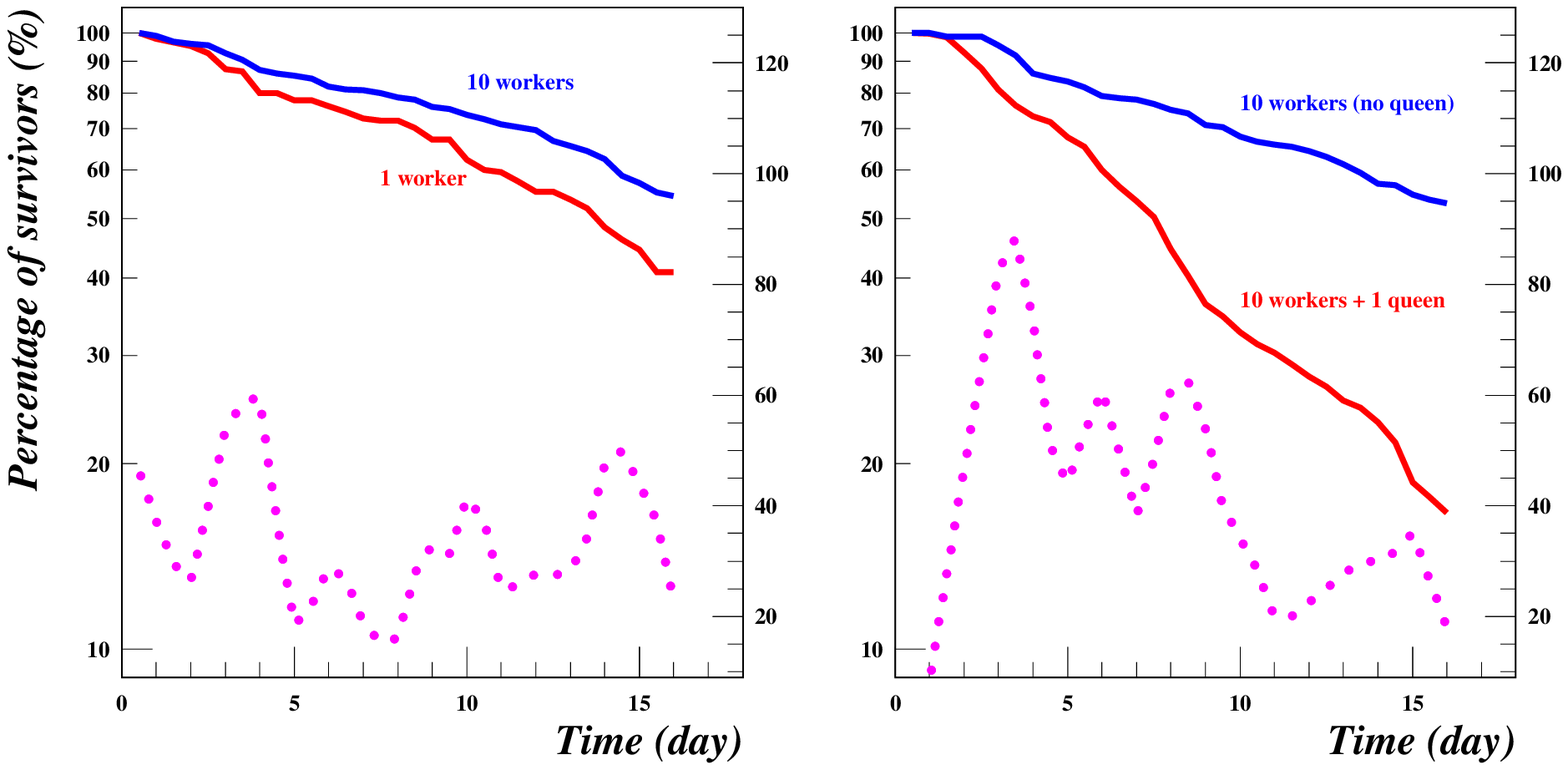}}
\qleg{Fig.\qhu 9\qhv Averaged survivorship curves for ants.}
{The graph gives averages of results obtained in similar
(but separate) experiments. The graph on the left-hand side is
an average over 4 experiments in which one compares groups
of 1 and 10 
whereas the graph on the right-hand
side is an average over 2 experiments in which one compares
groups of 10 with or without queen.
The dotted curves
give daily mortality rates expressed in deaths per day and per
1,000 individuals. The graphs suggest two conclusions:
(i) the survivorship curves are of type II (almost
constant mortality rate) (ii) Nevertheless in
the 4 days just after isolation
there is an upsurge in mortality that is significantly higher
than subsequent fluctuations.}
{Source: Results of individual experiments given in
Fig. 6 and Fig. 8.}
\end{figure}
%-------------------------------------------------
%

\qA{Survivorship curves for ant colonies (type II)}

So far we focused on the survival of individuals. 
What about the survival of {\it organizations}? 
From a systems science
point of view there is no crucial difference between 
individuals and organizations for both are a collection
of various elements held together by their common interactions.
However, it can be argued that individuals are systems that
are more integrated than organizations. For instance it 
is possible to split an organization into two parts
(as when one half of a beehive forms a swarm and leaves
the colony) whereas individuals cannot be divided into two
parts (except fairly rudimentary organisms, e.g. worms).

%%
%%%% PARALLELE ENTRE SURVIE DE COLONIES ET DE IT CORPORATIONS
\begin{figure}[htb]
\centerline{\psfig{width=8cm,figure=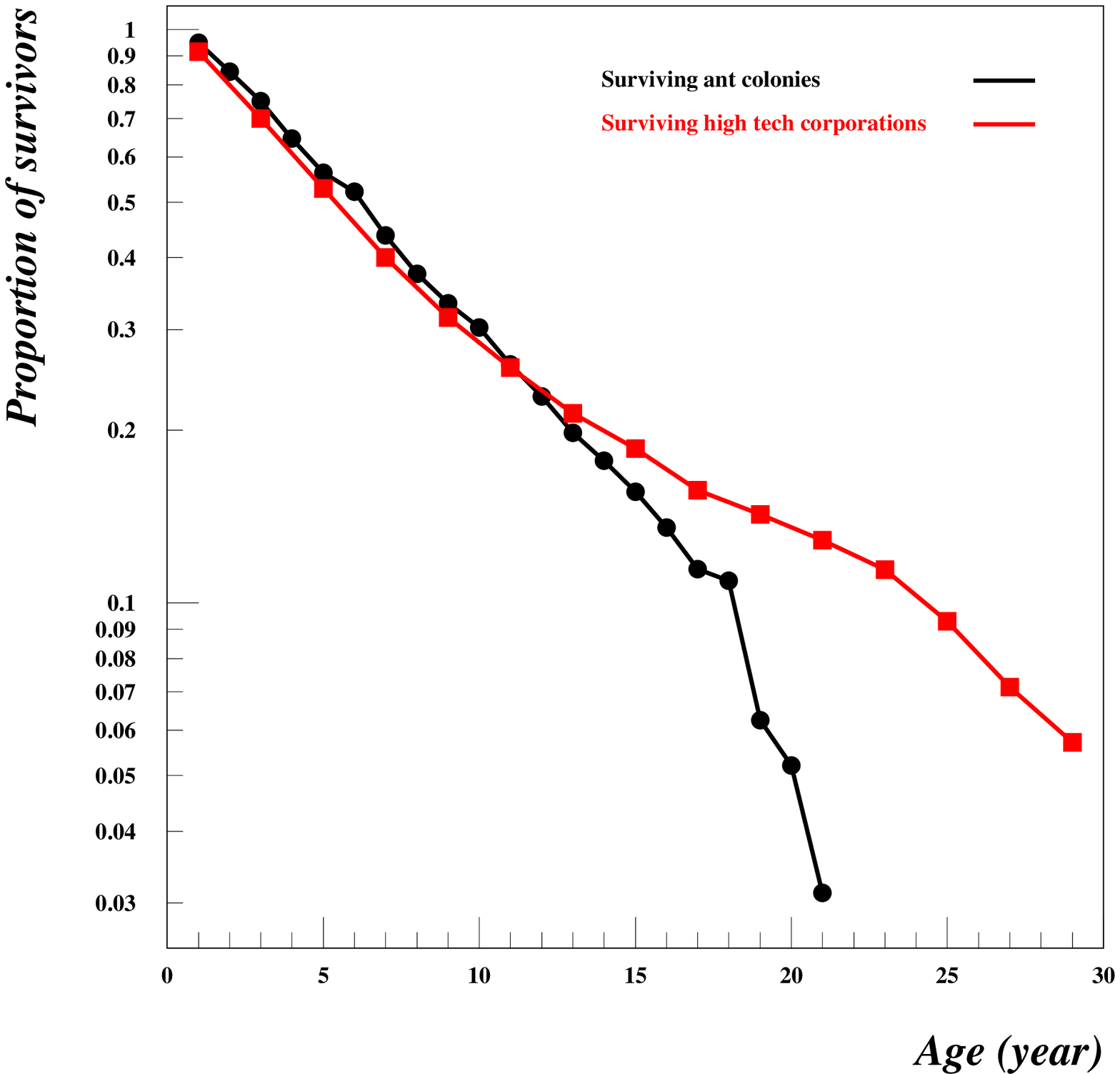}}
\qleg{Fig. \qhu 10\qhv Survivorship curves for ant
colonies and information technology corporations.}
{The curve for ant colonies is based on a sample
of 265 colonies. the curve for high-tech businesses
(in fact mostly information technology companies) is
based on a sample of 23,874 corporations in existence
between the years 1998 and 2009.}
{Sources: Ingram (2013, p. 7, Fig. 4a), 
Luo and Mann (2011, p. 9, chart 6). The authors would
like to thank Prof. Gordon for drawing their attention
on the data for ant colonies.}
\end{figure}
%-------------------------------------------------
%

Fig. 10 shows the survivorship curve for colonies 
of harvester ants, {\it Pogonomyrmex barbatus}, in a
place near the borderline between the US states of Arizona
and New Mexico. The curve is based on an annual census
of colonies performed by the team of Prof. Deborah Gordon
over a period of 28 years. In an area of about 10 hectares
($ 250 \times 375 $m) all new or extinct colonies were recorded
every summer. The colonies that
do not survive until they are 1 year old are not included.
The ``infant mortality'' in the first year can be estimated
separately
by comparing the numbers of reproductives and the numbers of 
new 1-year old colonies the following year. In this way, it was
found that fewer than 10\% of the new colonies survive 
through the first year.
\qpar

At first sight, the parallel made in Fig. 10 between
the survival of ant colonies on the one hand and high-tech
corporations on the other hand may seem surprising.
However, from a systems science perspective the two
processes are not so different. In both cases, the organization
has in principle the potential to develop but its growth
can be curtailed by a number of adverse conditions. In
that perspective the graph suggests a similarity in 
the balance between growth factors and 
growth obstacles, whether exogenous or endogenous.

\qA{Survivorship curves for humans and drosophila (type I)}

We already alluded to the fact that the human mortality
curve has the same shape as the mortality curve of
drosophila. If this is true for the mortality curve
it is of course even more true for the survivorship curve
which is basically the integral of the mortality curve.
\qpar

%%
%%%% PARALLELE ENTRE MORTALITE HOMMES ET DROSOPHILE
\begin{figure}[htb]
\centerline{\psfig{width=16cm,figure=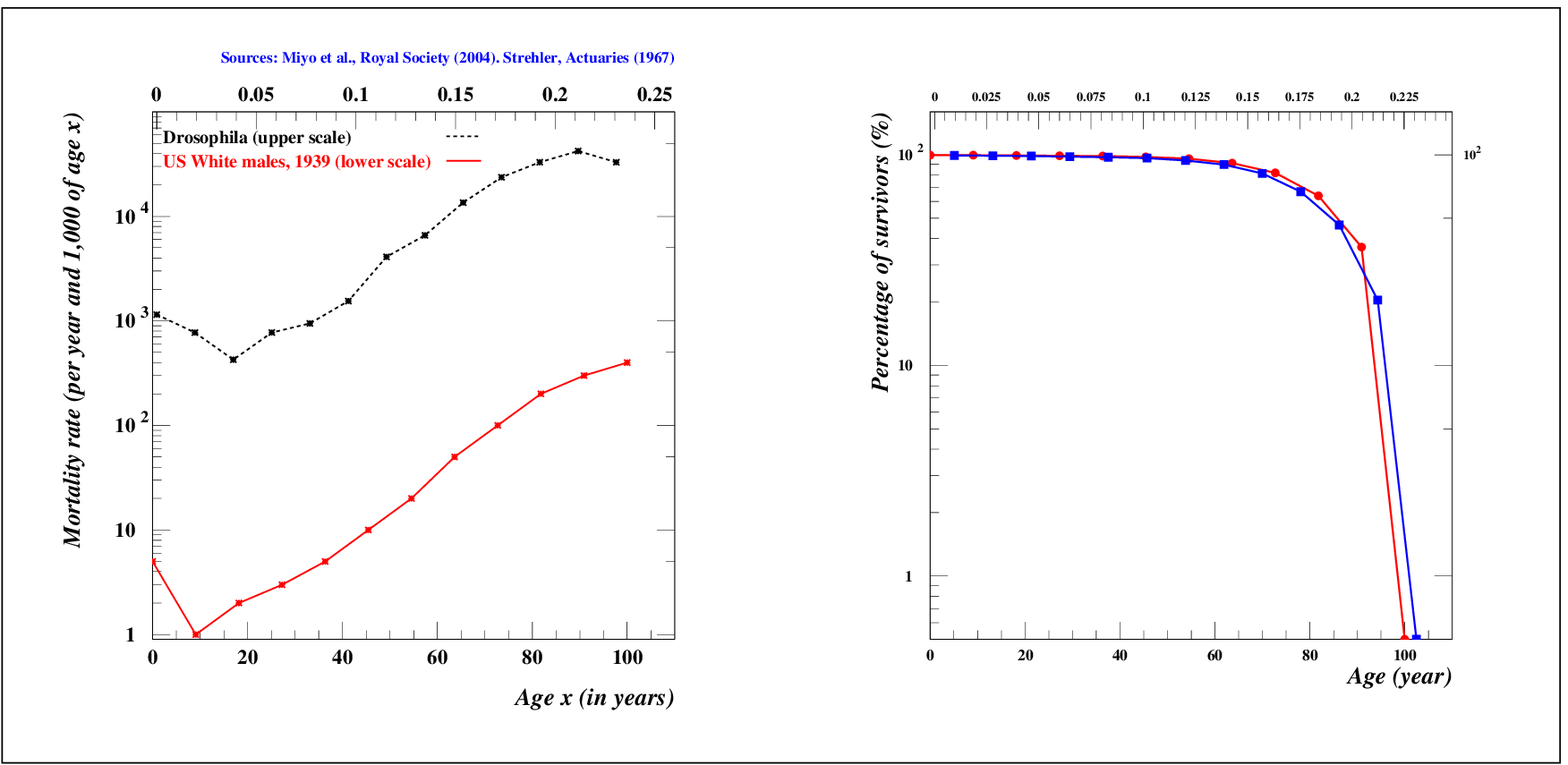}}
\qleg{Fig. \qhu 11\qhv Age-specific mortality rate for
humans versus drosophila.}
{The two curves on the left-hand side
display three common characteristics:
(i) High mortality shortly after birth (ii) exponential
growth (Gompertz law) in middle age (iii) leveling off 
in old age. The graph on the right-hand side shows the
corresponding survivorship curves which are the cumulative
sums of the mortality data.}
{Sources: Men: Strehler (1967); drosophila: Miyo and 
Charlesworth (2004)}
\end{figure}
%-------------------------------------------------
%
It is not clear whether the so-called $ r/K $ theory%
\qfoot{The two letters refer to the notation used to
write the Verhulst (also called ``logistic'') equation
namely:   
$$ { dN \over dt } = rN \left(1 - { N\over K }\right) $$
$ r $-species have many offsprings
while $ K $-species have few offsprings and live in fairly
dense habitat. Thus, drosophila would be an $ r $-species
while humans would be a $ K $-species.}
is able to provide a convincing explanation of
the parallelism observed in Fig. 11.

\qI{Transition dynamics from low to high mortality}

When a person becomes widowed does his (or her) mortality rate
immediately jump to the higher level which characterizes
this new situation or does it increase toward 
that level progressively with a time constant of a few years?
The same question can be asked for ants that are isolated
from their colony except that in this case we do not know 
very well the death rate before isolation that is to say 
in the colony. 

\qA{Three scenarios}
Three different scenarios are schematically displayed  
in Fig. 12. 
%%
%%%% TROIS SCENARIOS POUR LA TRANSITION 
\begin{figure}[htb]
\centerline{\psfig{width=8cm,figure=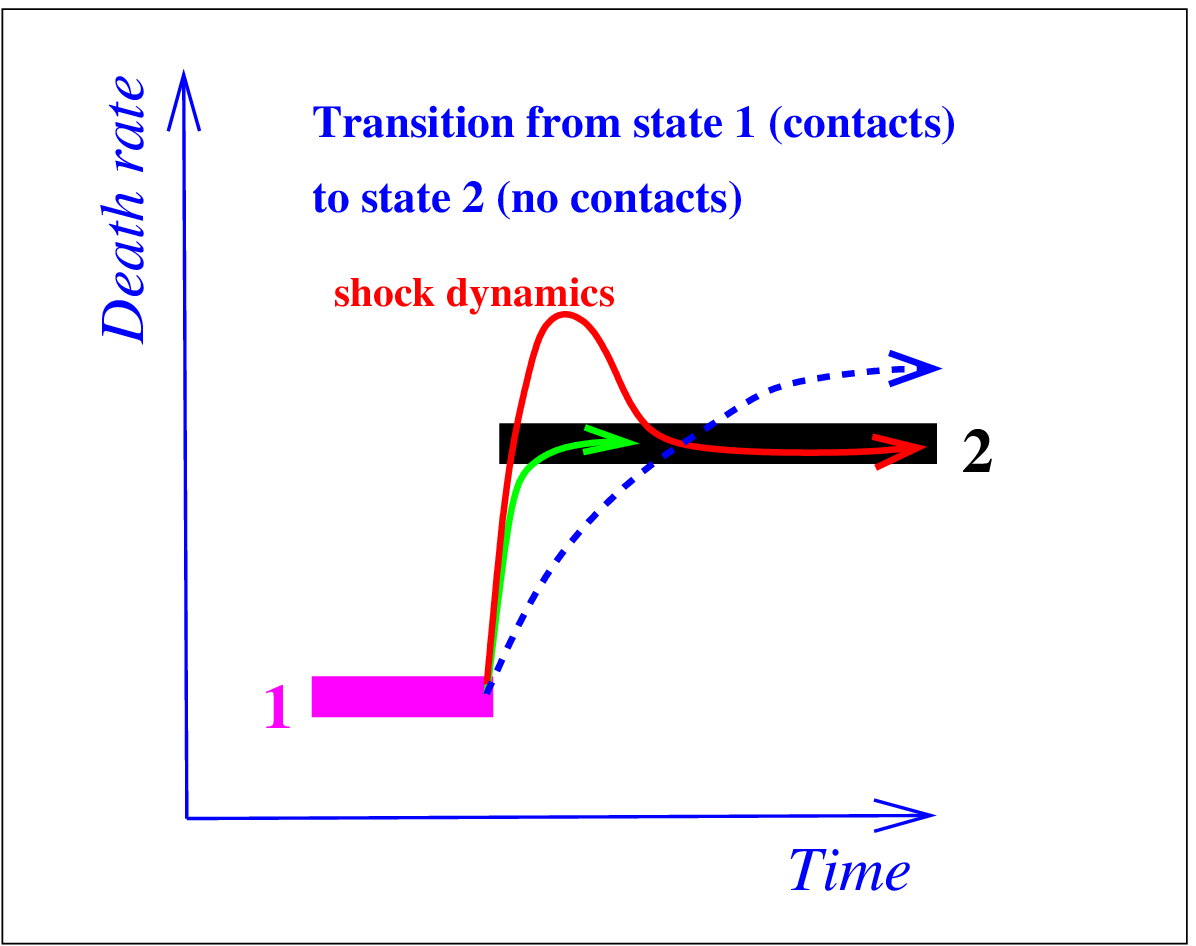}}
\qleg{Fig. \qhu 12\qhv Three scenarios for the transition
to a higher death rate.}
{The shock scenario is characterized by a death rate
that becomes temporarily higher than the stationary long-term
rate.}
{}
\end{figure}
%-------------------------------------------------

In addition to the abrupt change and smooth 
change scenarios
mentioned above there is a third one that we called
the shock scenario. It is characterized by a sharp
increase in death rate shortly after the links are
severed. Below we describe some pieces of evidence 
which support such a scenario. 

\qA{Evidence in favor of the shock scenario}
First we consider the case of ants. 
The graphs of average daily death rates
show an upsurge of deaths in the 3-4 days
after ants or fruit flies were put in isolation.
This effect is not due to the averaging process 
as can be seen by examining the graphs for separate
experiments. 
\qpar

The second piece if evidence comes from statistical
data about the transition to widowhood as summarized
in Table 3. It can be seen that during the first year 
after the marriage bond was broken the death rate
is higher than in subsequent years. The difference is
particularly clear for young widowers. 

%%-----------------------------------------------
\begin{table}[htb]

 \centerline{\bf Table 3: Excess mortality of widows in the years
following widowhood.}
\vskip 3mm
\hrule
\vskip 0.7mm
\hrule
\vskip 2mm

\color{black} 

$$ \matrix{
\hbox{Length} \hfill  & 35-44 & 35-44 & 45-54 & 45-54 & 55-64 & 55-64 
& 65-74 & 65-74 \cr 
\hbox{of time} \hfill & \hbox{M} & \hbox{F} &
\hbox{M} & \hbox{F} & \hbox{M} & \hbox{F} & \hbox{M} & \hbox{F}\cr
\hbox{after} & & & & & & & & \cr
\hbox{death of} & & & & & & & & \cr
\hbox{partner} & & & & & & & & \cr
\qtb
\hbox{[years]} & & & & & & & & \cr
\noalign{\hrule}
\qth 
\hfill \hbox{\bf 1} & 5.3 & 3.5 & 3.0 & 2.0 & 2.0 & 1.6 & 1.7 & 1.5\cr
\hfill \hbox{\bf 3} & 3.0 & 2.1 & 2.1 & 1.5 & 1.7 & 1.3 & 1.5  & 1.3 \cr
\qtb
\hfill >\hbox{\bf 3}& 2.7 & 1.7 & 2.0 & 1.4 & 1.6 & 1.3 & 1.3 & 1.2\cr
\noalign{\hrule}
} $$
\vskip 0.5mm
Notes: The table gives the ratio (mortality rate of widowed people)/
(mortality rate of married people of same sex and age); the data
are averages over 9 years, namely 1969-1974 and 1989-1991 (the
years refer to the years in which the people died). No data
were available under the age of 35. The shock effect
(that is to say a high excess mortality immediately after 
the death of the partner) is 
particularly clear for young men. 
\qL
{\it Source: Thierry (1999); primary source of the data:
INSEE (French National Statistical Office).}
\vskip 3mm
\hrule
\vskip 0.7mm
\hrule

\end{table}
%%-----------------------------------------------

\qI{Conclusion}
 First we summarize the main results, then we discuss
a possible agenda for further research.

\qA{Main results}
\qee{1} Although there are a few exceptions for separate
experiments, on average the isolated ants in groups of 10
have a lower mortality rate than those isolated alone.
A similar group effect is observed for married
versus non-married people.
\qee{1} The previous group effect is observed for social 
insects such as ants as well as for non-social insects such as the
{\it Bactrocera dorsalis} fruit fly.
\qee{3} However a major difference between ants and fruit
flies is the fact that if one excepts the burst
of deaths just after isolation,  
the death rate of isolated fruit
flies is at least 10 times smaller than the death rates of
ants. This conclusion holds for the groups of 10 as well
as for the singles. The experiments done by Grass\'e and
Chauvin show that for isolated bees the death rate is about
3 to 4 times higher than for ants. 
\qee{4} For both ants and fruit flies there is a shock effect
by which one understands that immediately after isolation
the death rate is much higher than during subsequent days.
This effect is particularly clear for fruits flies in which
case the initial death rate is about 5 times higher than
the subsequent death rate. This shock effect is also observed
for widowed people and especially for young widowers.
\qee{5} If queens are added to the groups of 10 fire ants
their mortality rate is multiplied by 2 or 3. This has been
observed repeatedly and without any exception.
\qee{6} In the standard classification of
survivorship curves, the curve of humans is of type I.
The same conclusion holds for drosophila. 
In the ant experiments the survivorship curves are
of type II. 
According to data for harvester ants from Deborah Gordon
the survival of newly founded ant colonies is also of type II.
The same conclusion holds for
newly founded information
technology corporations in the United States. 
In contrast, due to high initial death rate followed 
by a very low death rate,
the survivorship curve of isolated
{\it Bactrocera dorsalis} fruit flies is of Type III.\qL
Do standard survivorship types really provide useful clues 
for a better understanding? It does not seem so, one
must recognize.

\qA{Inference about the strength of social ties}

From the previous results is it possible to infer conclusions
regarding the strength of social interactions?
For living organisms without any interactions between one another
it would not make any difference to be separated from their
neighbors. Therefore one would not observe any shock effect
nor any death rate increase in groups of ``isolated''
individuals. As a matter of fact, the very notion of being
``isolated'' would have no meaning for such organisms.
On the contrary, for strongly interconnected organisms
it would make a big difference to be separated from their
neighbors. This is indeed confirmed by the data in Fig. 2 which
show that for low concentrations all cells die within 5-6 days.
\qpar
From these two extreme cases it is tempting to infer that
the increase in the death rate of isolated organisms is
an indicator of the strength of their interaction.
On this basis one can (tentatively) propose the following
ranking from weak to strong interactions.
\qee{i} Fruit flies ({\it Bactrocera dorsalis}) 
\qee{ii} Ants without queen
\qee{iii} Ants with queen 
\qee{iv} Married men over 50
\qee{v} Young married men under 30 
\qee{vi} Bees
\qee{vii} Cells such as those in Fig. 2.
\qpar
Needless to say, at this point this ranking is proposed rather
as a conjecture. If it can provide an incentive for further research
it will not be completely useless.

\qA{Agenda for further research}

We will consider successively short-term and long-term projects.
\qpar
One big uncertainty remains the average life duration
for ants in the colony. The estimate of 62 days for fire 
given above was a weighted average taking into account
differences in sizes but it was based on fairly imprecise 
data for each size class. In this connection it would
be interesting to see what happens when a big group 
comprising several hundred ants is isolated from the colony.
Will life expectancy in such a group approximate life
expectancy in the colony itself? One may get a clue by
increasing by steps the number $ n $ of ants in the group:
for instance, $ n=100, 1000, 10000 $. 
\qpar

In the experiments considered in this paper
one stumbling-block was the fact
that it takes at least one month to complete one of them.
It would be a major improvement to be able to reduce this time.
One possible method, namely Chauvin's starving
technique, was considered above (in the section about
experimental procedure) but turned out to be inappropriate
for ants or fruit flies. 
\qpar
We have seen that the initial shock which occurs in a 4-5 day
time interval after the beginning of isolation is a crucial
element of the whole process. Whereas the experimental procedure
(whether or not the dead ants are removed/replaced) may
affect the results over a period of one month, it seems plausible
that the initial shock would be less affected. If so, this would
provide a chance to do the main part of
the experiments in a shortened time. 
\qpar

The experiments considered so far provide more accurate
answers for several issues raised in the present paper.
Now, if one wishes
to broaden the exploration there is one question
which comes to mind immediately which is the following.
We have seen that the initial separation shock is observed
for social insects such as ants or bees as well as for
non-social insects such as fruit flies.
However, between social and non-social insects there is an
intermediate stage. In many species, e.g. the beetles
{\it Tenebrio molitor}, it is observed that, when put
together in a box, they will form a cluster (neither {\it Bactrocera
dorsalis} nor {\it Drosophila melanogaster} form clusters).
This means that Although such beetles apparently do not live in
colonies, they are nevertheless 
strongly attracted to one another (independently of sexual
attraction). 
Preliminary observation showed that a group of about 200
{\it Tenebrio molitor} does survive several months. It would 
be interesting to see what happens when subgroups of 1 or 10
individuals are separated from the main group.

\vskip 4mm
{\bf Acknowledgements} The authors are grateful
to Ms. Xin Liu, Yu Guo and  Zhaoqing Ding for their help
in the experiments. They also wish to thank Pr. Deborah
Gordon and Pr. Martin Raff for helpful suggestions.

\vskip 6mm
{\bf References} 
\qpar

\qparr
Arnold (G.) 1978: Les variations annuelles dans l'effet de groupe chez
l'abeille et l'origine de la mort pr\'ecoce des isol\'ees. 
[Changes in the group effect for bees and the explanation of the early
death of single bees.]
Insectes Sociaux 25: 39.

\qparr
Bertillon (L.-A.) 1872: Article ``Mariage'' in the
Dictionnaire Encyclop\'edique des Sciences M\'edicales,
2nd series, Vol. 5, p.7-52.

\qparr
Bertillon (L.-A.) 1879: Article ``France'' in the
Dictionnaire Encyclop\'edique des Sciences M\'edicales,
4th series, Vol. 5, p.403-584.

\qparr
Bose (E.) 1907: Resultate kalorimetrischer Studien. 
[Results from calorimetric investigations]
Zeitschrift
f\"ur Physikalische Chemie 58,585-624.

\qparr
Chauvin (R.) 1952: Sur le d\'eterminisme de l'effet de
groupe chez les abeilles. [Variability and invariance of
the group effect in bees].
Physiologia Comparata et Oecologia. An International
Journal of Comparative Physiology and Ecology 2,4,282-288.

\qparr
Chauvin (R.) 1973: Effet protecteur du groupement au cours 
du je\^une chez l'abeille.
[Group effect for the survival rate of starving bees.]
Comptes Rendus de l' Acad\'emie des  Sciences,
Paris [III] 276, 1479-1481.

\qparr
Chauvin (R.) 1981: La ``survivone'', substance qui induit la 
survie des abeilles isol\'ees." 
[``Survivone'', a substance which prevents the death of single bees.]
Insectes Sociaux 28,2,223-231.

\qparr
Chauvin (R.), Lafarge (J.-P.), Saligot (J.-P.): 1985: 
Action of a hormone, survivone, on the
mortality of isolated honeybees (group effect).
Apidologie: 16 77-87.

\qparr
%    (CD16,161)
Fielde (A.M.), Parker (G.H.) 1904: The reaction of ants
to material vibrations. Proceedings of the Academy of Natural
Sciences of Philadelphia, September 1904.

\qparr
%   (CD15,55)
Grass\'e (P.-P.), Chauvin (R.) 1944: L'effet de groupe dans la survie
des neutres dans les soci\'et\'es d'insectes 
[Collective effects in the
survival of workers in insect societies]. 
La Revue Scientifique Oct.-Nov., 461-464.

\qparr
Ingram (K.K.), Pilko (A.), Heer (J.), Gordon (D.M.) 2013: 
Colony life history and lifetime reproductive success of red
harvester ant colonies. 
Journal of Animal Ecology doi: 10.1111/1365-2656.12036

\qparr
Ishizaki (Y.), Voyvodic (J.T.), Burne (J.F.), Raff (M.C.) 1993:
Control of lens epithelial cell survival.
The Journal of Cell Biology 121,4,899-908 (May)

\qparr
Ishizaki (Y.), Burne (J.F.), Raff (M.C.) 1994: Autocrine
signals help chondrocytes to survive in culture.
The Journal of Cell Biology 126,4,1069-1077 (August)

\qparr
Legay (J.-M.), Pascal (M.) 1951: De l'effet de groupe chez le
ver \`a soie. [Group effect observation for silk worms.]
Comptes Rendus de l' Acad\'emie des  Sciences,
Paris 233,445-447.\qL
[The fertility of females is some 40\% higher in groups of 
250 individuals than in groups of 10.]

\qparr
Luo (T.), Mann (A.) 2011: High-tech business survivorship.
An analysis by organization type. 
Journal of Business and Economic Research 9,10.1-13.

\qparr
Merton (T.) 1956: On a barrier against insect pests.
Proceedings of the Royal Society of London, Series A,
Mathematical and Physical Sciences, 234,1197,218-220.

\qparr
Miyo (T.), Charlesworth (B.) 2004: Age-specific mortality rate
for Drosophila me\-lano\-gaster. Proceedings of Biological Sciences
of the the Royal Society 7 December, 271,\-1556,\-2517-2522.

\qparr
Raff (M.C.) 1998: Cell suicide for beginners. Nature 396, 12 November,
119-122.

\qparr
R\'eaumur (R.-A. Perchault de) 1928: Histoire des fourmis
[A natural history of ants]. Editions Paul Lechevalier.
[The publication was part No 11 of the 
``Encyclop\'edie Entomologique''. There may have been earlier
editions. Written around 1740, this was one of the earliest
known scientific study of the organization of ant colonies.]

\qparr
Roehner (B.M.) 2007: Driving forces in physical, biological and
socio-economic phenomena (in particular see
chapter 12 entitled ``Apoptosis'').
Cambridge University Press, Cambridge.

\qparr
Strehler (B.L.) 1967: Mortality trends and corrections.
Transactions of the Society of Actuaries 19,2,55,D428-D493.\qL
[Excerpt (p. D491-D492): ``If we plot the log of the probability
of death against age we get a Gompertz-type curve for virtually 
all populations; for smokers the curve is virtually advanced in age
so as to approximately double the normal mortality rate.
The average cigarette smoker can be considered as an individual who
is 7 year older than his chronological age. In like fashion,
exposure to ionizing radiation (in doses of 300 to 400 roentgens
total exposure) also increases age-specific death rate by 
a factor of about 2. If we were to eliminate all deaths due to
heart diseases and cancer, we would simply move down the Gompertz
curve by approximately 10 years.]

\qparr
Thierry (X.) 1999: Risques de mortalit\'e et de surmortalit\'e
au cours des 10 premi\`eres ann\'ees de veuvage. 
[Excess mortality in the 10-year period subsequently to widowhood.]
Population 54,2,177-204. 

\end{document}